\documentclass[11pt]{article}
\usepackage{amsfonts}
\usepackage{amssymb}
\usepackage{mathrsfs}
\usepackage{graphicx}
\usepackage{amsmath,amsthm,amssymb,amscd}
\usepackage[all]{xy}
\usepackage{subfig}
\usepackage{hyperref}
\usepackage{multirow}

\usepackage[dvips]{color}
\usepackage{amsfonts}
\usepackage{float}
\newcommand{\ca}{{\cal A}}
\newcommand{\cc}{{\cal C}}

\newcommand{\cn}{{\cal N}}

\newcommand{\cg}{{\cal G}}
\newcommand{\ch}{{\cal H}}

\newcommand{\co}{{\cal O}}

\newcommand{\cs}{{\cal S}}

\newcommand{\nn}{\nonumber}
\def\eqa{\begin{eqnarray}}
\def\eqae{\end{eqnarray}}
\def\eq{\begin{equation}}
\def\eqe{\end{equation}}
\def\be{\begin{equation}}
\def\ee{\end{equation}}
\def\bea{\begin{eqnarray}}
\def\eea{\end{eqnarray}}
\def\ba{\begin{array}}
\def\ea{\end{array}}
\def\bd{\begin{displaymath}}
\def\ed{\end{displaymath}}

\def\>{\rangle}
\def\<{\langle}
\def\a{\alpha}
\def\b{\beta}

\def\del{\delta}           
\def\e{\epsilon}           
\def\f{\phi}               
  
\def\g{\gamma}
\def\h{\eta}

\def\j{\psi}
                    
\def\l{\lambda}

\def\s{\sigma}

\def\F{\Phi}

\newcommand{\fn}{\footnotemark\footnotetext}

\def\NPB{Nucl.\ Phys.\ B}

\def\ZPC{Z.\ Phys.\ C}

\numberwithin{equation}{section}

\begin{document}

\begin{titlepage}
\begin{flushright}
MCTP 12-10 \\
PUPT 2416 \\
UCLA-12-TEP-102\\
\end{flushright}
\vspace{1cm}

\begin{center}
{\Large \bf One-loop renormalization and the S-matrix}\\[1cm]
Yu-tin Huang$^{a,b}$, David A. McGady$^c$ and Cheng Peng$^{d}$\\
~\\[5mm]

$^a${\it Department of Physics and Astronomy}\\
 {\it UCLA, Los Angeles}\\
 {\it CA 90095, USA}\\
~\\
$^b${\it School of Natural Sciences}\\
 {\it Institute for Advanced Study}\\
 {\it Princeton, NJ 08540, USA}\\
~\\
$^c${\it Department of Physics, Jadwin Hall}\\
{\it Princeton University}\\
{\it Princeton, NJ 08544, USA}\\
~\\

$^d${\it Michigan Center for Theoretical Physics}\\
{\it Department of Physics, University of Michigan}\\
{\it Ann Arbor, MI 48109, USA}\\
~\\

~\\
~\\
\today
~\\
~\\

\end{center}

\abstract{
In four-dimensional theories with massless particles, the one-loop amplitudes can be expressed in terms of a basis of scalar integrals and rational terms. 
Since the scalar bubble integrals are the only UV divergent integrals, the sum of the bubble coefficients captures the 1-loop UV
behavior. 
In particular, in a renormalizable theory the sum of the bubble coefficients equals the tree-level amplitude times a proportionality constant that is related to the one-loop beta function coefficient $\beta_0$. 
In this paper, we study how this proportionality is achieved from the viewpoint of on-shell amplitudes. For $n$-point MHV amplitude in (super) Yang-Mills theory, we demonstrate the existence of a hidden structure in each individual bubble coefficient which directly leads to systematic cancellations within the sum of them. The origin of this structure can be attributed to the collinear poles within a two-particle cut. Due to the cancellation, the one-loop beta function coefficient can be identified as a sum over the residues of unique collinear poles in particular two-particle cuts. Using CSW recursion relations, we verify the generality of this statement by reproducing the correct proportionality factor from such cuts for $n$-point split-helicity N$^{k}$MHV amplitudes.
}

\end{titlepage}

\tableofcontents
\newpage

\section{Introduction and summary of results}
\label{introduction}
In four spacetime dimensions, integral reduction techniques~\cite{PV,PV2,PV3} allow one to express one-loop gauge theory 
amplitudes in terms of rational functions and a basis of scalar integrals that includes boxes $I_4$, triangles $I_3$ and bubbles $I_2$~\cite{PV2, Bern:1994zx, Bern:1994cg}:
\begin{equation}
A^{\rm{1-loop}}=\sum\limits_i C_4^{i} I_4^{i}
  +\sum\limits_j C_3^{j} I_3^{j}
  +\sum\limits_k C_2^{k} I_2^{k}
  + \text{rationals} \ .
  \label{basis}
\end{equation}
Here the index $i$ ($j$ or $k$) labels the distinct integrals categorized by the set of momenta flowing into each corner of the box (triangle, or bubble).
In this basis, the scalar bubble integrals, $I_{2}^{i}$, are the only ultraviolet divergent integrals in four dimensions.
Moreover, the UV divergences of the bubble integrals take the universal form:
\eq
 I_{2}^{i}=\frac{1}{(4\pi)^2}\frac{1}{\e}+\mathcal{O}(1)
 \eqe
for all $i$.
Thus the \emph{sum of bubble coefficients} contains information on the ultraviolet behavior of the theory at one-loop.

In field theory, renormalizability requires that the ultraviolet divergences of the theory at one-loop can be removed by inserting a finite number of counterterms to the corresponding tree diagrams for the same process.
We can also understand this renormalizability from the amplitude point of view.
In terms of amplitudes, renormalizability implies that the ultraviolet divergence at one-loop must be proportional to the tree-amplitude.
As we will see in detail below, this proportionality between tree amplitudes and the bubble coefficients, which encapsulate UV behavior of the theory, in renormalizable theories is cleanly illustrated in pure-scalar QFTs.
In $\f^{4}$ theory, the bubble coefficient of the 4-point one-loop amplitude evaluates to the 4-point tree amplitude $\parbox[c]{.5cm}{\includegraphics[width=1.8cm]{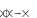}}$\hspace{1.38cm}.
However, in $\f^{5}$ theory, the bubble coefficient of the simplest 1-loop amplitude evaluates to a 6-point amplitude $\parbox[c]{.5cm}{\includegraphics[width=1.8cm]{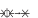}}$\hspace{1.38cm}.
Similarly, this new 6-point amplitude will generate higher-point amplitudes at higher loops, which is the trademark of a non-renormalizable theory.
This observation connects renormalizability with the 1-loop bubble coefficient: in a renormalizable theory, the \emph{sum of bubble coefficients} is proportional to the tree amplitude
\begin{equation}\label{bbprop}
\cc_2\equiv\sum_iC^{i}_2\propto A^{\rm tree}\,.
\end{equation}
where the sum $i$ runs over all distinct bubble cuts, and we use the calligraphic $\cc_2$ to denote the \emph{sum} of the bubble coefficients.
This proportionality relation takes a very simple form in (super) Yang-Mills theory with all external lines being gluons \cite{divs,QCDbeta,SimplestQFT} (see~\cite{ehp} for detailed discussion)
\eq
\cc_2=-\beta_0A^{\rm tree}_n\,,\hspace{12mm} \b_{0}=- \bigg( \frac{11}{3} n_v - \frac{2}{3} n_f - \frac{1}{6} n_s \bigg)\,.
\label{constraint}
\eqe
where $\beta_0$ is the coefficient of the one-loop beta function and $n_{v}, n_{f}, n_{s}$ are numbers of gauge bosons, fermions and scalars respectively. From the amplitude point of view, eq.~(\ref{constraint}) appears to be a miraculous result as each individual bubble coefficient is now a complicated rational function of Lorentz invariants.
For example, it is shown in \cite{Lal:2009gn, ehp} that for the helicity amplitude $A_{4}(1^{+}2^{-}3^{+}4^{-})$ in $\cn$-fold super Yang-Mills  theory, the bubble coefficients of the two cuts are:
 \eqa
& & C_{2}^{(23,41)}=-(\cn-4) \frac{\<12\>\<34\>}{\<13\>\<24\>} A_{4}^{\text{tree}}(1^{+}2^{-}3^{+}4^{-}) \, ,  \,{\rm and} \, \nonumber\\
& & C_{2}^{(12,34)}=-(\cn-4) \frac{\<14\>\<23\>}{\<13\>\<24\>}A_{4}^{\text{tree}}(1^{+}2^{-}3^{+}4^{-}) \, . \nonumber
\eqae
However the sum of these two bubble coefficients is exactly proportional to the tree amplitude: $C_{2}^{(23,41)}+C_{2}^{(12,34)}=-(\cn-4) A_{4}^{\text{tree}}(1^{+}2^{-}3^{+}4^{-})=-\b_{0}
A_{4}^{\text{tree}}(1^{+}2^{-}3^{+}4^{-})$ by the Schouten identity.
For an arbitrary $n$-point amplitude, eq.~(\ref{constraint}) implies cancellation among a large number of these rational functions, in the end yielding a simple constant multiplying $A_n^{\rm tree}$.
The fact that the proportionality in eq.~\eqref{constraint} holds for any renormalizable theory, hints at possible hidden structures in the sum of the bubble coefficients. Note that for gauge theories with non-adjoint matter fields, the individual bubble coefficients will also depend on higher order Casimir invariants~\cite{Lal:2009gn}. Renormalizability then requires all the higher order invariants to cancel in the sum, leaving behind only the quadratic Casimir $tr_R(T^aT^b)$. In this paper, we seek to partially expose hidden structure of the bubble coefficients that leads to the proportionality to the tree amplitude.

Following~\cite{Forde:2007mi, SimplestQFT}, we extract the bubble coefficient by identifying it as the contribution from the pole at infinity in the complex $z$-plane of a BCFW-deformation~\cite{BCFW} of the two internal momenta in the two-particle cut, where the complex deformation is introduced on the internal momenta. We begin with scalar theories as a warm up. Here the contributions to bubble coefficients are tractable using Feynman diagrams in the two-particle cut. For scalar $\phi^n$ theories, we demonstrate that the bubble coefficient only receives contributions from one-loop diagrams that have exactly two loop-propagators. For each diagram, the contribution is proportional to a tree diagram with a new $2(n-2)$-point interaction vertex. Renormalizability requires $n=2(n-2)$, so this implies the familiar result, $n = 4$.

Feynman diagrams become intractable in gauge theories and it is simpler to use helicity amplitudes in the cut. In (super) Yang-Mills theory, we study general MHV $n$-point amplitudes and find that for each 2-particle cut, the bubble coefficient can be separated into four separate terms.
 Each term stems from the four distinct singularities which appear as the loop momenta become collinear to one of the adjacent external legs, indicated in Fig.~\ref{CCP}~(a).
 We show that these singularities localize the Lorentz invariant phase space ($d{\rm LIPS}$-) integral to residues at four separate poles. Once given in this form, we find: 
\begin{figure}
\begin{center}
\includegraphics{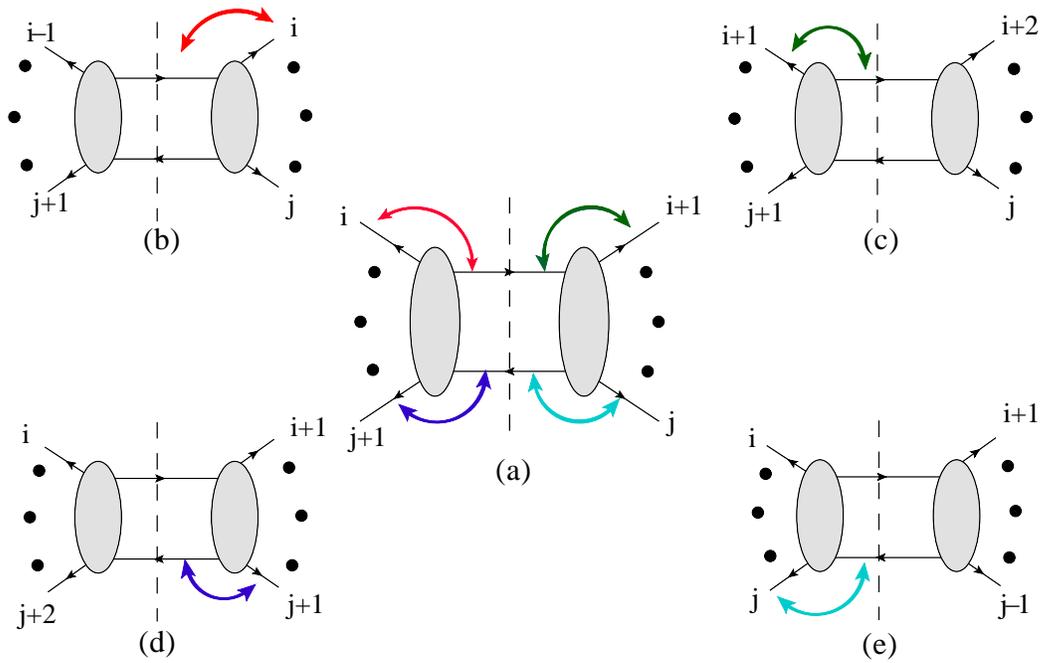}
 \caption{A schematic representation of the cancellation of common collinear poles (CCP).
 The bubble coefficient of the cut in figure (a), receives contributions from the four collinear poles indicated by colored arrows.
 Each collinear pole is also present in the corresponding adjacent cut indicated in figures (b), (c), (d), and (e) respectively.
 In the sum of bubble coefficients such contributions cancel in pairs. }
\label{CCP}
\end{center}
\end{figure}

\begin{figure}
\begin{center}
\includegraphics[width=3in]{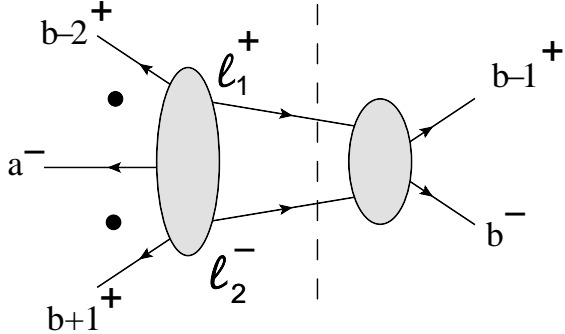}
\caption{The terminal channels that gives non-trivial contribution to the sum of bubble coefficients. Note the helicity configurations of the loop legs of the $n$-point tree amplitude is identical with the two external legs on the 4-point tree amplitude in the cut. }
\label{TargedIn}
\end{center}
\end{figure}

\begin{itemize}
\item 
\indent \textit{For each collinear residue in a generic cut, there is a residue in the adjacent cut that has the same form but with opposite sign.}
When we sum over all channels, residues stemming from common collinear poles (CCP) in \emph{adjacent} channels cancel pair-wise, as indicated in Fig.~\ref{CCP}.
The sum therefore telescopes to four unique poles that come from four distinct ``terminal cuts''. Here we define a terminal cut as the two particle cut which contains at least one 4-point tree amplitude on one side of the cut. The poles in interest correspond to the point in the phase-space where the two on-shell loop momenta become collinear with the two external scattering states in the 4-point sub-amplitude. We will refer to these poles as ``terminal poles''. 

\item 
\indent Focusing on the terminal poles we find that their contributions to the bubble coefficients are non-trivial only if the helicity configuration of the particles crossing the cut is ``preserved", i.e. the loop helicity configuration is the same as the external lines on the 4-point tree amplitude as shown in Fig.~\ref{TargedIn}. Thus the beta function of (super) Yang-Mills theory is given by the residues of the helicity conserving terminal poles. 
\end{itemize}
For MHV amplitudes, we show that for a two non-vanishing terminal poles whose residues are identical and equal to $ 11/6 A_{n}^{\text{tree}}$.
Summing the two then gives the desired result, $\cc_2=11/3 A_n^{\rm tree}$ for the pure Yang-Mills theory, in agreement with eq.~(\ref{constraint}).
The relation \eqref{constraint} is also derived in the super Yang-Mills theory where $\cc_{2}=-(\mathcal{N}-4) \mathcal{A}_n^{\rm tree}$ for ${\cn = 1, 2}$.

For general N$^k$MHV split-helicity amplitudes in pure Yang-Mills theory, we also show that the residue of each helicity conserving terminal pole give $11/6 \, A_{n}^{\text{tree}}$. We demonstrate this by using the CSW~\cite{csw} representation for the N$^k$MHV tree amplitudes appearing in the two-particle cut. The fact that these terminal cuts give the correct proportionality factor indicates that these are indeed the only non-trivial contributions to the sum of bubble coefficients. This also hints at systematic cancellation in the sum of bubble coefficients should be a property of Yang-Mills amplitude for general helicity configuration. We give supporting evidence by using the collinear splitting function to show that the residues of CCP in a two particle cut for generic gauge theories are indeed identical with opposite signs.

Our paper is organized as follows.
In section~\ref{scalarQFT}, we compute the bubble coefficients for theories of self-interacting scalar fields, and rederive the well-known renormalizability conditions.
 We proceed to analyze (super) Yang-Mills theories with emphasis on the cancellation of common collinear poles (CCP) in section~\ref{gluonQFT}. We will use super Yang-Mills MHV amplitudes as the simplest demonstration of such cancellation. Similar results occur for MHV amplitudes in Yang-Mills as well. In section~\ref{CCCP}, we give an argument for the cancelation of CCP for generic external helicity configurations by showing, using splitting functions of the tree amplitude in the cut, that the residue of  collinear poles of the entire cut is indeed shared with an adjacent channel. We present further evidence in section~\ref{NMHV_Sect} by explicitly proving that the forward limit poles for split-helicity N$^k$MHV amplitudes indeed give the complete RHS of eq.~(\ref{constraint}), implying complete cancellation of all other contributions.

\section{Bubble coefficients in scalar field theories}\label{scalarQFT}

\begin{figure}
\begin{center}
\includegraphics[width=2in]{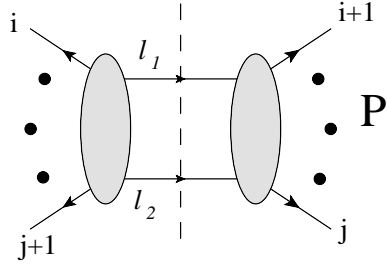}
\caption{A two-particle cut for a generic $n$-point amplitude. }
\label{New}
\end{center}
\end{figure}

As a toy model, we consider scalar theories with single interaction vertex $\a_k\phi^k$ in this section. It was shown in~\cite{SimplestQFT}, following previous work in~\cite{Forde:2007mi}, that the bubble coefficient for a given two-particle cut can be calculated as:
\begin{equation}\label{bubble1}
  C^\text{(i,j)}_\text{2} =
  \frac{1}{(2\pi i)^2}
  \int d \text{LIPS}[l_1,l_2]
  \int_\mathcal{C}\frac{dz}{z} ~\widehat{S}^\text{(i,j)}_{n}
  \end{equation}
where $(i,j)$ indicates the momentum channel $P=p_{i+1}+\cdots+p_j$ of the cut as shown in Fig.~\ref{New},
$\widehat{S}^\text{(i,j)}_{n}=\hat{A}_L^\text{tree}(|\hat{l}_1\>,|\hat{l}_2])\, \hat{A}_R^\text{tree}(|\hat{l}_1\>,|\hat{l}_2])$, and
$d$LIPS$=d^4l_1d^4l_2 \,\delta^{(+)}(l_1^2)\,\delta^{(+)}(l_2^2)\,\delta^{4}(l_1+l_2-P)$.
Here $\hat{A}^{\text{tree}}_{L,R}$ in \eqref{bubble1} are the amplitudes on either side of the cut; hats in \eqref{bubble1} indicate a BCFW shift \cite{BCFW} of the two cut loop momenta:
\begin{equation}
\widehat{l}_1(z) = l_1  + q   z, \qquad \widehat{l}_2(z) = l_2  - q  z\,, \qquad \text{with}~~q\cdot q =q\cdot l_{1}=q\cdot l_{2}=0\,.\label{bcfwshift}
\end{equation}
We integrate the shift parameter $z$ along a contour $\mathcal{C}$ that goes around infinity, which evaluates to the residue at the $z=\infty$ pole of the integrand.\fn{ The BCFW shifts of the two-particle cut allows one to explore all possible on-shell realizations of a double-cut for a given set of kinematics. The presence of finite-$z$ poles indicates the existence of additional propagators, which are the contributions of box or triangle integrals to the double-cut.  The contribution from the bubble integrals then correspond to poles at $z=\infty$, hence the choice of contour. }

\begin{figure}
\begin{center}
\includegraphics{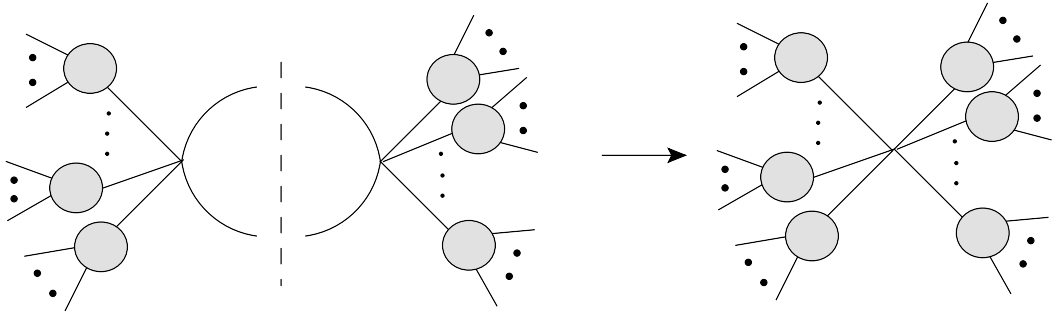}
\caption{For pure $\phi^k$ theory, the only one-loop diagrams that gives non-trivial contribution to the bubble integrals are those with only two loop propagator. The contribution to the bubble coefficient is simply the product of the tree diagrams on both side of the cut, connected by a new $2(k-2)$ vertex.}
\label{Collapse}
\end{center}
\end{figure}

In a scalar theory, the only $z$ dependence in BCFW-shifted tree-amplitudes comes from propagators which depend on one of the two loop momenta. Under BCFW-deformations, propagators of this type scales as $\sim1/z$ for large-$z$.  Diagrams containing such propagators die-off as $1/z$ or faster. The only non-vanishing contribution to the bubble coefficient comes from diagrams with the two shifted lines on the same vertex \cite{Feng:2009ei}. In this case there is neither $z$-dependence nor dependence on $l_{1}$, or $l_{2}$ in the double-cut and \eqref{bubble1} evaluates to
\begin{eqnarray}
C^\text{(i,j)}_\text{2} &=&
  -\frac{1}{2\pi i}
   \int d \text{LIPS}~A^{\text{tree}}_{L}A^{\text{tree}}_{R}=A^{\text{tree}}_{L}A^{\text{tree}}_{R}\,,
\label{shrink}
\end{eqnarray}
where $A^{\text{tree}}_{L,R}$ are the unshifted amplitudes on either side of the cut as in Fig. \ref{Collapse}, and we have used $\frac{1}{2\pi i} \int d \text{LIPS} (1)=-1$ (Appendix~\ref{Holomorphic}). The bubble coefficient \eqref{bbprop} is a sum over all cuts.

Before the general $n$-point analysis, we first consider explicit examples at 4- and 6-points. For simplicity, we consider these examples with  {\bf \emph{color-ordered amplitudes}}, where the scalars transform under the \emph{adjoint representation} of some gauge group. We will switch to the  \emph{non-color-ordered amplitudes} later for the general analysis.

At 4-point, the tree amplitude is $A^{\rm tree}_{4} = \a_4$. There are two cuts of the 1-loop 4-point amplitudes, namely the $s$ and $t$ channels. Then \eqref{shrink} gives
\begin{eqnarray}
\cc_{2}  &=& A^{\rm tree}_{4}(1,2, \hat{l}_{1}, \hat{l}_{2})\times A^{\rm tree}_{4}(-\hat{l}_{2}, -\hat{l}_{1},3,4) + A^{\rm tree}_{4}(4,1, \hat{l}_{1}, \hat{l}_{2})\times A^{\rm tree}_{4}(-\hat{l}_{2}, -\hat{l}_{1},2,3)\nonumber \\
 &=&  \a_4^2+\a_{4}^{2}= 2 \a_4 A_4^{\rm tree}\,.
\end{eqnarray}

At 6-point, the tree amplitudes is
$A^{\rm tree}_{6}=\frac{1}{S_{123}}+\frac{1}{S_{612}}+\frac{1}{S_{561}}$.
There are six cuts with two shifted lines sitting on one vertex, whose bubble coefficients are listed in the following table
\begin{center}
\begin{tabular}{|c|c|c|c|c|c|c|}
  \hline
  Channel  & $\scriptsize \frac{\a_{4}^{2}}{S_{123}}$& $\scriptsize \frac{\a_{4}^{2}}{S_{234}}$ &$\scriptsize\frac{\a_{4}^{2}}{S_{345}}$ &$\scriptsize\frac{\a_{4}^{2}}{S_{456}}$&$\scriptsize\frac{\a_{4}^{2}}{S_{561}}$&$\scriptsize\frac{\a_{4}^{2}}{S_{612}}$\\[0.8mm]\hline\hline
  \scriptsize $(12|3456)$ & ~ & ~ & 1 & 1 & ~ &  ~\\\hline
   \scriptsize $(23|4561)$ & ~ & ~ & ~ & 1 & 1 &  ~\\\hline
    \scriptsize $(34|5612)$ & ~ & ~ & ~ & ~ & 1 &  1\\\hline
     \scriptsize $(45|6123)$ & 1 & ~ & ~ & ~ & ~ & 1\\\hline
      \scriptsize $(56|1234)$ & 1 & 1 & ~ & ~ & ~ &  ~\\\hline
       \scriptsize $(61|2345)$ & ~ & 1 & 1 & ~ & ~ &  ~\\\hline  \hline
        $\cc_{2}$ & 2 & 2 & 2 & 2 & 2 &  2\\\hline
\end{tabular}\label{sixpoint}
\end{center}
From this table, we see
\begin{equation}
\cc_{2}=2 \a_{4}^{2}\bigg(\frac{1}{S_{123}}+\frac{1}{S_{234}}+\frac{1}{S_{345}}+\frac{1}{S_{456}}+\frac{1}{S_{561}}+\frac{1}{S_{612}}\bigg)=4\a_{4}A^{\rm tree}_{6}\,.
\end{equation}
For {\bf\emph{non-color-ordered amplitudes}}, 
\begin{equation}\label{treebubb0}
  \cc_2= \frac{3}{2}\a_4 (n-2) A_n^{\rm tree} =\frac{ (n-2)}{2} \b_{0} A_n^{\rm tree}\,, \qquad \b_{0}=3\a_{4}\, .
\end{equation}

\begin{figure}
\begin{center}
\includegraphics[width=4in]{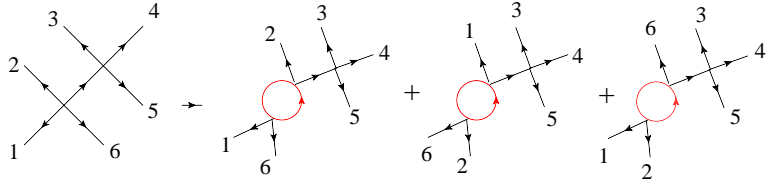}
\caption{For any given tree-diagram in the $\phi^4$ theory, each vertex can be blown up into 4-point one-loop subdiagrams in three distinct ways, while preserving the tree graph propagators. Each case contributes a factor of $\a_4$ times the original tree diagram to the bubble coefficient. In this figure we show the example of 6-point amplitude.}
\label{Blowup}
\end{center}
\end{figure}

The above result can also be understood by simply noting the fact that the contribution to the bubble coefficient of each one-loop diagram, is in fact proportional to a tree diagram. More precisely:
\begin{itemize}
  \item  From \eqref{shrink}, the bubble coefficient for each cut equals the product of the two cut amplitudes. This value coincides with the tree diagram obtained by replacing the loop with a 4-point contact vertex multiplied by a factor of $\alpha_4$.
  \item  Reverse the above statement. Each 1-loop diagram with a non-zero bubble coefficient can be obtained by taking a tree diagram and ``blowing up'' one of the interaction vertices into a one-loop sub-diagram. Taking all distinct tree diagrams and ``blowing up'' one vertex at a time, produces all relevant one-loop diagrams.
  \item There are three different ways of ``blowing up'' each distinct interaction vertex, as indicated in Fig~\ref{Blowup}. Essentially these are the $s,t,u$ channels for a given one-loop four-point amplitude. Each of these one-loop diagrams gives an identical contribution to the overall bubble coefficient.
  \end{itemize}

In $\f^4$-theory, each $n$-point tree-diagram has $(n-2)/2$ vertices, hence there are $3(n-2)/2$ one-loop diagrams that contribute to the bubble coefficients. This gives $3 \times (n-2)/2$ copies of the original tree diagram. This proves \eqref{treebubb0}.

Further, we can re-derive the standard renormalizability criterion for 4-dimensional scalar field theories. The result~\eqref{shrink} is true for any $\a_{k}\f^{k}$ interaction. It suggests that we can think of the bubble coefficient as shrinking the loop to generate a new $2(k-2)$ vertex, which is illustrated in Fig.~\ref{Collapse}. Thus we see that the sum of bubble coefficients for the $n$-point amplitude is a sum of $n$-point tree-diagrams that are constructed from the original $k$-point contact vertex as well as one $2(k-2)$-vertex. As discussed in the introduction, the renormalizability requires that the bubble coefficient is proportional to the tree amplitude. Since the tree-level amplitudes are built from $k$-point vertices only, the proportionality between bubble coefficient and tree amplitude \eqref{bbprop} requires
\eqa\label{consistency}
2(k-2) = k\;\;\rightarrow\;\; k=4 \,.
\eqae
This reproduces the known result of renormalizability from the power-counting arguments of local perturbative QFT in four dimensions.

Above we studied scalar theories with just a single real scalar field.
It is useful to now consider a model with two real scalars $\phi_1$ and $\phi_2$ and an interaction $\phi_1^2 \phi_2^2$.
The sum of bubble coefficients for the 4-point amplitude $A_4(\phi_1,\phi_1,\phi_1,\phi_1)$ will be proportional to tree amplitudes with a single $\phi_1^4$-vertex.
Similarly for $\phi_2$.
So even if $\phi_1^4$ and $\phi_2^4$ were not part of the original theory, they were generated at loop-level.
Since $A_4^\text{tree}(\phi_1,\phi_1,\phi_1,\phi_1)$ vanishes in the original theory, one might now object to the statement that for renormalizable theories (such as this one, of course) the sum of bubble coefficients $\mathcal{C}_2$ is proportional to the tree amplitude.
However, the point here is whether the ``new" interactions give rise to an infinite tower of other new interactions or not.
The former would not be a renormalizable theory.
An infinite tower would be generated in the case of $\phi^5$ while for our example here, $\phi_1^4$ and $\phi_2^4$ do not generate any new interactions.
So when we say that in a renormalizable theory, the sum of bubble coefficients $\mathcal{C}_2$ is proportional to the tree amplitude, then we regard the tree level theory as the renormalizable theory with generic couplings.
In that case, the two-scalar theory does include $\phi_1^4$ and $\phi_2^4$ and we simply learn from the bubble coefficients
that these couplings are renormalized.

We can do a similar analysis to the Yukawa theory with complex scalars. For \emph{pure scalar amplitudes}, Feynman diagrams are tractable and  eq.~\eqref{bbprop} renormalizes the coupling constant of the $\phi^2\phi^{*2}$ interaction in the action. For \emph{amplitudes with external fermions}, complications arise due to the $z$-dependence of external line factors of the fermions and we use helicity amplitudes. As an example we show that eq.~\eqref{bbprop}  does not generate the non-renormalizable $\phi^2\psi^2$ or $\psi^4$ interaction terms, as expected. We show the detailed analysis in Appendix \ref{YukawaQFT}.

\section{Bubble coefficients for MHV (super) Yang-Mills amplitude}\label{gluonQFT}

When we consider the (super) Yang-Mills theory, the proportionality between the sum of bubble coefficients and the tree amplitude becomes extremely non-trivial. Here, individual bubble coefficients are generically complicated rational functions of spinor inner products as illustrated for the  $\<\F_{1}\Psi_{2}\F_{3}\Psi_{4}\>$ case in the introduction. In general, only after summing all the bubble coefficients and repeated use of Schouten identities, will the result reduce to a simple constant times $\mathcal{A}^{\rm tree}_n$.
Thus from the amplitude point of view, this proportionality is a rather miraculous result.

\begin{figure}
\begin{center}
\includegraphics[width=4in]{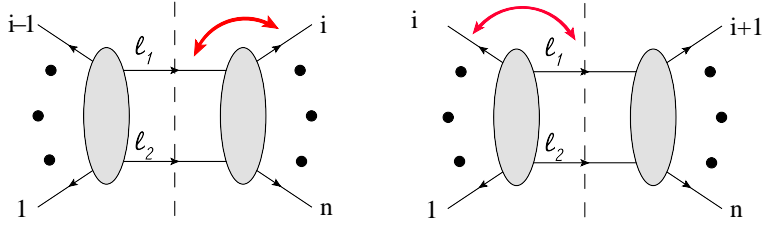}
\caption{ An illustration of the cancellation between adjacent channels. The contribution to the bubble coefficient coming from the $d{\rm LIPS}$ integral evaluated around the collinear pole $\langle l_1i\rangle\rightarrow0$, indicated by the (red) arrows,  of the two diagrams cancels as indicated in eq.~\eqref{cancel}.}
\label{AdjacentCuts}
\end{center}
\end{figure}

In this section we show that the cancellation is in fact systematic. To see this, we show that for MHV amplitudes, the $d{\rm LIPS}$ integration will be localized by the collinear poles of the tree amplitude on both sides of the two-particle cut.
For a generic cut, there are four distinct collinear poles involving the loop legs, each of which is also present in an adjacent cut, as illustrated in Fig.~\ref{AdjacentCuts}.
It can be shown that the residues of these two adjacent cuts on their common collinear pole, $\<\l, i\> \to 0$, are exactly equal and with opposite sign.
By separating the bubble coefficient into four different terms, corresponding to contributions from four different  poles, the cancellation between common collinear poles (CCP) in the sum of bubble coefficients is manifest.

\begin{figure}
\begin{center}
\includegraphics[width=2in]{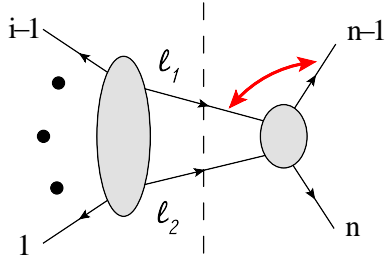}
\caption{ The ``terminal'' pole that contributes to the bubble coefficient. Such poles appear in the two particle cuts that have two legs on one side of the cut and one of the legs has to be a minus helicity.  Note that, at this point in phase-space, $l_1 = -p_{n-1}$ and $l_2 = -p_n$.}
\label{Terminal}
\end{center}
\end{figure}

Cancellation stops at ``terminal cuts" where a 4-point tree and an $n$-point tree appear on opposites sides of the cut. The uncancelled terms in these terminal cuts correspond to the residues of collinear poles where the two loop-momenta become collinear with the external momenta of the two external legs on the 4-point amplitude, as illustrated in Fig.~\ref{Terminal}.
Explicitly, for \emph{adjoint} fields (vectors, fermions and scalars), we see the sum of these ``terminal poles'' is 
\eqa
-\b_0 A_n^{\rm tree}(1^+\ldots a^-,\ldots,b^- \ldots n^+) = \bigg(\frac{11}{3} n_v - \frac{2}{3} n_f - \frac{1}{6} n_s \bigg) \frac{\<a, b \>^4}{\<1, 2 \> ... \< i, i+1 \> ... \<n, 1 \>} \, ,
\eqae
for MHV amplitudes with $n-2$ positive-helicity gluons and negative-helicity gluons $a$ and $b$ \cite{ParkeTaylor}.  In the following, we will demonstrate this for $n$-point MHV amplitudes in ${\cn = 1, 2}$ super Yang-Mills theory.  This systematic cancellation is also present for pure Yang-Mills MHV amplitudes (explicitly shown in appendix~\ref{NAMHV}).

Before going further, we pause to note an important distinguishing feature between the bubble coefficients in scalar QFT and in (S)YM.  Specifically, the proportionality constant in (super) Yang-Mills is independent of the number of external legs: it is just -$\beta_0$, the coefficient of the one-loop beta function.  To see this note that for (super)Yang-Mills theory, there are diagrams with one-loop bubbles in the external legs. These massless bubbles do not appear in eq.~(\ref{basis}), as they are set to zero in dimensions regularization, reflecting the cancellation between collinear IR and UV divergences.
However, when one is only considering the pure UV divergence of the amplitude, one must take into the account the existence of the UV divergences in the external bubble diagrams, which are simply the same as that of the infrared divergences in the external bubbles, but with a relative minus sign. Thus we have
\begin{eqnarray}
  \nonumber
 A_n|_{ \text{UV-div.}}
  &=&
  \big(\sum C_\text{bubble} I_\text{bubble}\big)_\text{UV}
  + \text{UV}_\text{ext.\,bubbles}\\
  &=& \big(\sum C_\text{bubble} I_\text{bubble}\big)_\text{UV}
  - \text{IR}_\text{ext.\,bubbles}\,.~
\end{eqnarray}
For $n$-gluon 1-loop amplitudes, the collinear IR divergences take the form~\cite{divs}
\bea
  \label{coll}
   \text{$\text{IR}$:}~~~~~~~~~
   A^\text{1-loop}_{n,\text{collinear}}
  ~=~ -\frac{g^2}{(4\pi)^2} \frac{1}{\epsilon} ~ \frac{n}{2}\, \beta_0 A^\text{tree}_{n} \, .
\eea
At leading order in $\epsilon \to 0$,
the UV divergence is \cite{divs}
\bea
  \label{UV}
   \text{$\text{UV}$:}~~~~~~~~~
 A^\text{1-loop}_{n,\text{UV}}
  ~=~+\frac{g^2}{(4\pi)^2} \frac{1}{\epsilon} ~ \bigg(\frac{n-2}{2}\bigg)\, \beta_0 A^\text{tree}_{n} \, .
\eea
Thus the bubble coefficients (total UV divergence) in \emph{purely} gluonic one-loop amplitudes are
\eq
\sum_{i}C_{2}^{i}=A^\text{1-loop}_{n,\text{UV}}+A^\text{1-loop}_{n,\text{collinear}}=-\beta_0\,A_{n}^\text{tree} = \frac{11}{3}\,A_n^{\rm tree} \, .
\eqe

At one loop, $\phi^4$ scalar field theory lacks these collinear divergences on external legs, and no UV/IR mixing occurs, hence pure scalar bubble coefficients scale with  $\frac{n-2}{2}$, the number of interaction vertices.

\subsection{Extracting bubble coefficients in (${\cn = 0, 1, 2}$ super) Yang-Mills}
The bubble coefficient for a given two-particle cut of a one-loop (S)YM amplitude is computed in essentially the same way as for scalar field theory.  However, as emphasized in the introduction, unlike the case for scalar QFT extracting this through Feynman diagrams is rather intractable.  Roughly in YM this is because BCFW shifts of the two internal on-shell gluon lines in the double-cut introduces $z$-dependence in local interaction vertices \emph{and} polarization vectors.   These difficulties are only amplified in ($\cn \neq 0$) SYM.

It is more efficient to directly express the LH- and RH- amplitudes as entire on-shell objects through use of the spinor-helicity formalism.  Here the $d{\rm LIPS}$ integration over allowed on-shell momenta is conveniently converted into an integration over spinor variables which automatically solve the delta functions,
\eq
\int d^4l_1d^4l_2 \,\delta^{(+)}(l_1^2)\,\delta^{(+)}(l_2^2)\,\delta^{4}(l_1+l_2-P)g(|l_1\>,|l_2\>)= \int_{\tilde{\l}=\bar{\l}} P^2 \frac{\<\l, d\l \> [\tilde{\l}, d\tilde{\l}]}{\< \l | P | \tilde{\l}]^2} g(|\l\>,P|\tilde\l])\,,
\label{Basics}
\eqe
where we have identified $|l_1\>=|\l\>$, $|l_2\>=P|\tilde\l]$, and $\int_{\tilde{\l}=\bar{\l}}$ indicates we are integrating over the real contour (real momenta). The $\widehat{S}^\text{(i,j)}_{n}$ in \eqref{bubble1} takes the form
\be\label{symc2def}
  \widehat{S}^{(i,j)}_{n}= \widehat{\cs}^{(i,j)}_{n,\,0}\equiv
  \sum_\text{state sum}  \hat{A}_L^\text{tree}(|\hat{l}_1\>,|\hat{l}_2])\,
  \hat{A}_R^\text{tree}(|\hat{l}_1\>,|\hat{l}_2])
   \,,
\ee
in the Yang-Mills theory.  Note that to fully integrate out the bubble coefficients' dependence on the internal lines, we sum over all possible states in the loop.

Further, extraction of simple bubble coefficients is aided by on-shell SUSY.\fn{The calculations for the simplest bubble coefficients are simpler in $\cn = 1, 2$ SYM than in YM. To see this, compare non-adjacent MHV bubble computations in SYM (subsection~\ref{SYM_MHV}) and in YM (appendix~\ref{NAMHV}).}  Here amplitudes and state sums are promoted to superamplitudes and Grassmann integrals
\begin{equation}
  \widehat{S}^{(i,j)}_{n}=\hat{\cs}^{(i,j)}_{n,\,\cn}~\equiv~\sum_{\sigma} \int d^\cn\h_{l_1} d^\cn\h_{l_2}~\hat{\ca}_{L\sigma}^\text{tree}\,\hat{\ca}_{R\bar{\sigma}}^\text{tree}\,, \qquad\qquad \cn=1,2\,,
\end{equation}
where $\sigma$ labels the different pairs of multiplets that the loop legs, $l_1$ and $l_2$, belong to. Following~\cite{ehp}, on-shell states are encoded into two separate on-shell superfields, $\F$ and $\Psi$, that contain states in the `positive' and `negative-helicity' sectors. In this language, $\{ \sigma \} = \{ (\Phi,\Psi), (\Psi,\Phi), (\Phi,\Phi), (\Psi,\Psi)\}$.  The $\bar{\sigma}$ is the conjugate configuration of $\s$.

Crucially, to preserve SUSY the \emph{bosonic} BCFW shift~\eqref{bcfwshift} must be combined with a \emph{fermionic} shift of the Grassmann variables $\eta^a$~\cite{Ferber:1977qx, SimplestQFT}
 \begin{subequations}\label{supershift}
 \begin{align}
|\widehat{l_1} (z) \rangle &= | l_1 \rangle + z | l_2 \rangle,\qquad\qquad |\widehat{l_2} (z) ] = |l_2 ] - z | l_1 ]\,,   \label{supershiftb}\\
 \hat{\eta}_{l_{2}a} &= \eta_{l_{2}a} + z \eta_{l_{1}a}, ~\qquad\qquad a=1,\ldots,\cn \, . \label{supershiftf}
 \end{align}
\end{subequations}
Note the bosonic shift~\eqref{supershiftb} is identical to the shift~\eqref{bcfwshift}, when cast in terms of the spinor-helicity variables; it is referred to as an $[l_2,\,l_1\>$-shift.

Combined super-shifts~\eqref{supershift}, of any tree amplitude of the $\cn=4$ SYM fall-off as $1/z$ for large-$z$.
In (S)YM theory with $\cn=0,1,2$ supersymmetry, it was shown~\cite{ehp} that the super-BCFW shifts $[\F,\F\>$, $[\Psi,\F\>$ and $[\Psi,\Psi\>$ fall off as $1/z$ at large $z$ while the $[\F,\Psi\>$ super-shift grows as $z^{3-\cn}$ for large-$z$. For $\cn=0$ pure Yang-Mills, this reduces to the familiar observation that for shifts $[-,-\>, \ [-, + \>$, and $[+, + \>$ the amplitudes fall off as $1/z$, while the $[+,-\>$ shifts grow as $z^3$~\cite{BCFW,SpinLorentz}.

Carrying out the $z$ integral gives
\begin{eqnarray}\label{bubble2}
  C^{(i,j)}_\text{2} &=&
  -\frac{1}{2\pi i}
   \int d \text{LIPS}[l_1,l_2]~
    \Big[ \widehat{\cs}^{(i,j)}_{n,\,\cn}\Big]_{O(1)\text{ as }z\to\infty} \,, \, \,  \cn=0,1,2\,,
\end{eqnarray}
where $\Big[ \widehat{\cs}^{(i,j)}_{n,\,\cn}\Big]_{O(1)\text{ as }z\to\infty}$ is the residue of $\widehat{\cs}^{(i,j)}_{n,\,\cn}$ at the $z\to \infty$ pole.

\begin{figure}[t]
  \centering
  \subfloat[Cuts with $\s \in \{(\Phi,\Phi), (\Psi,\Psi)\}$ die-off, and have no pole, as $z \to \infty$.]{\includegraphics[scale=0.9]{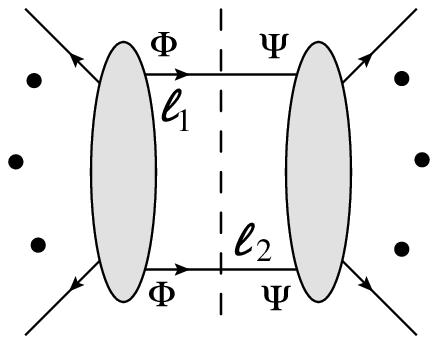}
\label{ppmm}} \quad\quad\quad\quad
  \subfloat[Cuts with $\s \in \{(\Phi,\Psi), (\Psi,\Phi)\}$ have a pole at $z \to \infty$ ($\cn = 0, 1, 2$).]{\includegraphics[scale=0.9]{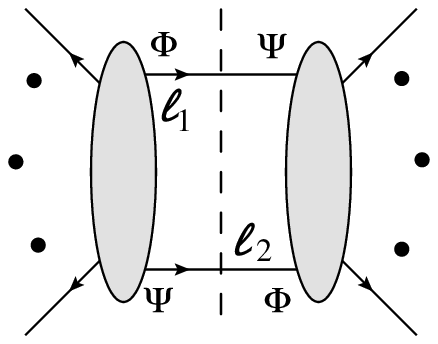}
  \label{pmmp}}
  \caption{Illustration of two internal helicity configurations of the cut loop momenta. }
  \label{2hel}
\end{figure}

Double-cuts with internal states $\s \in \{ (\Phi, \Phi), (\Psi, \Psi) \}$, shown in cut $(a)$ of Fig.~\ref{2hel}, scale as
\bea
   \widehat{\cs}^\text{\,[Cut (a)]}_{n,\,\cn}
   ~\sim~ \frac{1}{z} \times \frac{1}{z}
   ~\sim~ \frac{1}{z^2}~~~\text{as}~~~z \to \infty \,.
   \label{ALARz}
\eea
Cuts of this type do not contribute to the bubble coefficient. On the other hand, cuts with internal states $\sigma\in \{(\Phi,\Psi),(\Psi,\Phi)\}$, such as cut $(b)$ in Fig.~\ref{2hel}, always involve a shift that acts as $[\Psi,\Phi\>$ on one sub-amplitude and as $[\Phi,\Psi\>$ on the other. This gives
\bea
  \label{scutN}
      \widehat{\cs}^\text{\,[Cut (b)]}_{n,\,\cn}
   ~\sim~ \frac{1}{z} \times z^{3-\cn}
   ~\sim~ z^{2-\cn}
   ~~~\text{as}~~~z \to \infty \,.
\eea
Note that \eqref{scutN} indicates that there can be non-vanishing $O(1)$-terms and hence contributions to the sum of the bubble coefficients in $\cn=0,1,2$ SYM but  not in the $\cn=4$ SYM theory. This is consistent with the known non-vanishing 1-loop $\b$-functions in $\cn=0,1,2$ SYM  theories as well as the UV-finiteness of $\cn=4$ SYM.

We can also investigate the contribution to the bubble coefficients from particular states crossing the two-particle cut. These contributions can be projected out by acting with appropriate Grassmann integrations/derivatives on the above superamplitudes. By analyzing the large-$z$ behavior of the integration measure, as shown in \cite{ehp}, we get the large-$z$ behavior of the bubble coefficient of certain internal states. The implication for QCD with one flavor of fermions can simply be deduced from $\mathcal{N}=1$ super Yang-Mills, where there are no scalars in the multiplet. For example, a simple computation shows that a double-cut of an internal negative-helicity gluino and an internal negative helicity gluon exiting (entering) one of the two tree-amplitudes does not contribute to the bubble coefficients in $\cn=1$ SYM theory. Then following \cite{Dixon:2010ik}, this result is also true in QCD with one flavor of fermions. Note that the difference between fundamental and adjoint fermions is irrelevant for this analysis since we are interested in color-ordered amplitudes and the large-$z$ behavior of the cut holds for individual internal helicity configurations and not the sum.

\subsection{MHV bubble coefficients in ${\cn = 1, 2}$ super Yang-Mills theory}\label{SYM_MHV}

It was shown in ref.~\cite{ehp}, that for the MHV amplitudes in $\mathcal{N}=1,2$ super Yang-Mills theory, the $\mathcal{O}(z^0)$ part of the BCFW-shifted two-particle cut $\widehat{\cal S}_{\{a,b\},\,\cn}^{(i,j)}$ is given by:
  \begin{eqnarray}
\widehat{\cal S}_{\{a,b\},\,\cn}^{(i,j)}\bigg|_{{\cal O}(z^0)} =({\cal N}-4) {\cal A}_n^{\text{tree}}\frac{\<i, i+1\> \<j, j+1\>}{\<a, b\>^2}\frac{\< a, \l \>^2 \< b, \l \>^2}
{\< j, \l \> \< j+1, \l \> \< i, \l \> \< i+1, \l \>}  ,
\label{S}
\end{eqnarray}
where $\{a,b\}$ indicates the positions of the two sets of external negative-helicity states (within the $\Psi$ multiplets).
We have set  $|l_2\>=|\l\>$. Since $\widehat{{\cal S}}_{\{a,b\},\,\cn}^{(i, j)}$ is purely holomorphic in $\l$, we can straightforwardly use eq.~\eqref{derivative} to rewrite the $d{\rm LIPS}$ integral \eqref{Basics} as a total derivative, and the bubble coefficient is given as:
\begin{figure}
    \centering
     \includegraphics[width=2.5in]{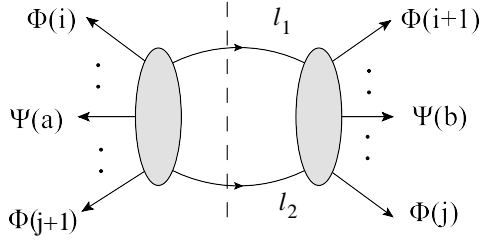}
      \caption{The two-particle cut that gives $\widehat{\cal S}_{\{a,b\},\,\cn}^{(i,j)}$.}
       \label{GeneralCut}
\end{figure}
\eqa
\nonumber
C^{(i,j)}_{2\;\{a,b\}}&=&\frac{-1}{2 \pi i} \oint_{\tilde{\l} =\overline{\l}}
P_{i+1,j}^2 \frac{\< \l, d\l \> [\tilde{\l}, d\tilde{\l}]}{\< \l | P_{i+1,j} | \tilde{\l}]^2} ~\widehat{\cal S}_{\{a,b\},\,\cn}^{(i,j)}\bigg|_{{\cal O}(z^0)}  \\
&=&\frac{1}{2 \pi i}\oint_{\tilde{\l} = \overline{\l}} \< \l, d\l \> d\tilde{\l}_{\dot{\a}}
\frac{\partial}{\partial \tilde{\l}_{\dot{\a}}}
\bigg[
\frac{[ \tilde{\l} | P_{i+1,j} | \a\> }{\< \l | P_{i+1,j}  | \tilde{\l}] \< \l, \a \>}~ \widehat{\cal S}_{\{a,b\},\,\cn}^{(i,j)}\bigg|_{{\cal O}(z^0)}
\bigg]  \,,\label{FF}
\eqae
where $|\a\>$ is a reference spinor. In this section, for convenience, we label the bubble coefficients ${ C}^{(i,j)}_{2\;\{a,b\}}$ in the same way as the two-particle cut $\widehat{\cal S}_{\{a,b\},\,\cn}^{(i,j)}$. There are two kinds of poles inside the total derivative, the four collinear poles of $\widehat{\cal S}_{\{a,b\},\,\cn}^{(i,j)}$ in eq.~(\ref{S}) and the spurious pole $1/\< \l, \a \>$. The spurious pole can be simply removed by the $\< a, \l \>^{2}\< b, \l \>^{2}$ factor in the numerator of eq.~(\ref{S}) if we choose the auxiliary spinor $| \a \>$ to be $|a\>$ or $|b\>$. Thus with this choice of reference spinor, the contributions to the bubble coefficient come solely from the collinear poles in $\widehat{\cal S}_{\{a,b\},\,\cn}^{(i,j)}|_{{\cal O}(z^0)}$.

From eq.~(\ref{S}) we see that there are four collinear poles in $\widehat{\cal S}_{\{a,b\},\,\cn}^{(i,j)}\big|_{{\cal O}(z^0)}$, each corresponding to $\lambda$ becoming collinear with the adjacent external lines of the cut. Careful readers might find this puzzling, as the MHV tree amplitudes on both side of the cut only have collinear poles of the form $\<l_1,i\>$, $\<l_1,i+1\>$, $\<l_2,j\>$ and $\<l_2,j+1\>$. Recalling that here $|\l\>=|l_2\>$, one would instead expect collinear poles of the form, $[\l|P_{i+1,j}|i\>$, $[\l|P_{i+1,j}|i+1\>$, $\<\l,j\>$ and $\<\l,j+1\>$. The resolution is that eq.~(\ref{S}) is obtained by shifting $\<l_1, i\>\rightarrow\<l_1, i\>+z\<l_2,i\>$ and expanding around $z\rightarrow\infty$, thus introducing the $\<l_2,i\>$ poles:
\begin{equation}
\frac{1}{\<l_1(z), i\>}\bigg|_{z\rightarrow\infty}=\frac{1}{z\<l_2,i\>}+\co\bigg(\frac{1}{z^{2}}\bigg) \, .
\end{equation}
Since these poles originated from $\<l_1(z),i\>$, we will abuse the terminology, as well as the figures, and still refer to them as collinear poles.\footnote{In fact, this is not as much of an abuse as it may seem. Note that evaluating the pole at $z\rightarrow\infty$ is equivalent to evaluating the pole at the origin minus the poles at finite $z$. The former would be a true collinear pole, while the latter would be a collinear pole with shifted $l_1$.\label{footnote}}

\begin{figure}
    \centering
     \includegraphics[width=2.5in]{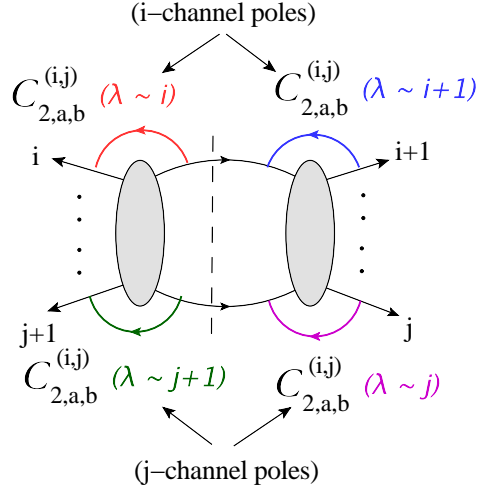}
      \caption{A graphical representation of eq.~(\ref{Mess}). The bubble coefficient of a given channel is separated into four terms, each having a different collinear pole as the origin of the holomorphic anomaly that gives a non-zero $d{\rm LIPS}$ integral. The 4-contributions can be grouped into two channels, the $i$- and the $j$-channel.}
       \label{MessFig}
\end{figure}

To better track the contributions of the collinear poles, we rewrite the integrand as follows:
\eqa
\nonumber
\widehat{\cal S}_{\{a,b\},\,\cn}^{(i,j)}\bigg|_{{\cal O}(z^0)} &=&({\cal N}-4) {\cal A}_{n}^{\text{tree}}\frac{\< a, \l \> \< b, \l \>^2\<i, i+1\>}{\<a, b\>^2\< i, \l \> \< i+1, \l \>}\left(\frac{\< a, j+1 \>}
{\< j+1, \l \>}-\frac{\< a,j \>}
{\< j, \l \> }\right)\\
&=&({\cal N}-4) {\cal A}_n^{\text{tree}}\frac{\< a, \l \> \< b, \l \>^2 \<j, j+1\>}{\<a, b\>^2 \< j, \l \> \< j+1, \l \> }\left(\frac{\< a, i+1 \>}
{ \< i+1, \l \>}-\frac{\< a, i \>}{ \< i, \l \>}\right)\,,\label{factor}
\eqae
where the two equivalent representations focus on different adjacent collinear poles in the parentheses. The representation in eq.~(\ref{factor}) allows us to compute the bubble coefficient in a manner that manifests the relation between collinear poles in adjacent channels. With auxiliary spinor $|\alpha\>$ in eq.~(\ref{FF}) chosen to be $|a\>$, the bubble coefficient is
\eqa
\nonumber{C}^{(i,j)}_{2\,\{a,b\}} = \ {C}^{(i,j)}_{2\,\{a,b\}}(\l \sim j+1)&+& {C}^{(i,j)}_{2\,\{a,b\}}(\l \sim j)\ + \ {C}^{(i,j)}_{2\,\{a,b\}}(\l \sim i+1)\ + \  {C}^{(i,j)}_{2\,\{a,b\}}(\l \sim i)\,.
\eqae
Here we have used $(\l \sim j)$ to indicate the contribution from the collinear pole $\<\l,j\>$. For convenience, we will refer to $(\l \sim j)$ and $(\l \sim j+1)$ collinear poles as ``$j$-channel poles'', and $(\l \sim i)$ and $(\l \sim i+1)$ poles as ``$i$-channel poles''. A graphical illustration of eq.~(\ref{Mess}) is given in Fig.~\ref{MessFig}

Before proceeding, we point out a very important observation.  Comparing the first line of eq.~(\ref{factor}) for $\widehat{\cal S}_{\{a,b\},\,\cn}^{(i,j)}|_{{\cal O}(z^0)}$ with that for $\widehat{\cal S}_{\{a,b\},\,\cn}^{(i,j-1)}|_{{\cal O}(z^0)}$,
$$\widehat{\cal S}_{a,b}^{(i,j-1)}\bigg|_{{\cal O}(z^0)} =({\cal N}-4) {\cal A}_{n}^{\text{tree}}\frac{\< a, \l \> \< b, \l \>^2\<i, i+1\>}{\<a, b\>^2\< i, \l \> \< i+1, \l \>}\left(\frac{\< a, j \>}{\< j, \l \>}-\frac{\< a,j-1 \>}{\< j-1, \l \> }\right)\,,$$
we immediately see that \textit{terms containing the common collinear pole of the two adjacent cuts, i.e. $1/\<j,\l \>$, are exactly the same but, crucially, have opposite signs}. This applies to all of the other terms in eq.~(\ref{factor}), each having a counterpart in the adjacent channel, as illustrated in Fig.~\ref{CCP}.

At this point, one is tempted to conclude that the contribution to the bubble coefficient from common collinear channels cancels. However there is one subtlety. In eq.~(\ref{FF}), besides $\widehat{\cal S}_{\{a,b\},\,\cn}^{(i,j)}|_{{\cal O}(z^0)}$, there is an extra factor in the total derivative that depends on the total momentum of the two-particle cut, $P_{i+1,j}$, which will be distinct for the adjacent cuts. Luckily these distinct factors become identical on the common collinear pole:
\eq
\frac{[ \tilde{\l} | P_{i+1,j} | a\> }{\< \l | P_{i+1,j}  | \tilde{\l}] \< \l, a \>}\bigg|_{\<\l,j\>=[\tilde{\l},j]=0}=\frac{[ \tilde{\l} | P_{i+1,j-1} | a\> }{\< \l | P_{i+1,j-1}  | \tilde{\l}] \< \l, a \>}\bigg|_{\<\l,j\>=[\tilde{\l},j]=0}\,,
\eqe
where $|_{\<\l,j\>=[\tilde{\l},j]=0}$ indicates that the loop momentum is evaluated in the limit where it is collinear with $j$.\footnote{Since the contour of the $d{\rm LIPS}$ integral is taken to be real, $\tilde{\lambda}=\bar{\lambda}$, the collinear pole $1/\<\l,j\>$ freezes the loop momenta to satisfy $\<\l,j\>=[\tilde{\l},j]=0$. } Because the extra factors are identical on the common collinear pole (CCP), we now conclude that the contribution of the CCP to the bubble coefficient indeed cancels between adjacent channels. This can also be concretely checked against the result from the direct evaluation of the $d$LIPS integral: \fn{The explicit evaluation of integrals in eq.~(\ref{Basics}) using the holomorphic anomaly \cite{csw,SimplestQFT, ehp} is reviewed in appendix~\ref{Holomorphic}.}
\eqa
\ {C}^{(i,j-1)}_{2\,\{a,b\}}(\l \sim j)&=&-({\cal N} - 4) {\cal A}_n^{\rm tree} \frac{\< i, i+1 \>}{\< a, b \>^2}\frac{\< a | P_{i+1,j-1} | j]}{ \< j+1 | P_{i+1,j-1} | j]}
\frac{\< a, j \> \< b, j \>^2}{\< i, j \> \< i+1, j \>}\,,
 \nonumber\\
{C}^{(i,j)}_{2\,\{a,b\}}(\l \sim j)&=&({\cal N} - 4) {\cal A}_n^{\rm tree} \frac{\< i, i+1 \>}{\< a, b \>^2}\frac{\< a | P_{i+1,j} | j ]}{\< j | P_{i+1,j} | j ]}
\frac{\< a, j \>  \< b, j \>^2}{\< i, j \> \< i+1, j \>}\,. \nn
\eqae
Adding these two equations, we find
\eqa
\nonumber&&\hspace{-2mm}{ C}^{(i,j)}_{2\,\{a,b\}}(\l \sim j)+{ C}^{(i,j-1)}_{2\,\{a,b\}}(\l \sim j)\\ \label{cancel}
&=&\hspace{-2mm}(4-{\cal N}) {\cal A}_n^{\rm tree} \frac{\< i, i+1 \>}{\< a, b \>^2}
\frac{\< a, j \>  \< b, j \>^2}{\< i, j \> \< i+1, j \>}\bigg[-\frac{\< a | P_{i+1,j} | j ]}{\< j | P_{i+1,j} | j ]}
+\frac{\< a | P_{i+1,j-1} | j ]}{ \< j | P_{i+1,j-1} | j ]} \bigg]=0\,,\label{Mess}
\eqae
thus verifying our claim.

Since the four collinear poles for a generic two-particle cut are shared by four different adjacent channels as shown in Fig.~\ref{CCP}, this immediately leads to the result that although the bubble coefficient for a generic two-particle cut is given by complicated rational functions, as shown in eq.~(\ref{Mess}),  \textit{in summing over all two-particle cuts there is a pairwise cancellation of CCP, and thus a majority of bubble coefficients do not contribute to the final result. }

\begin{figure}[htbp]
\begin{center}
\includegraphics{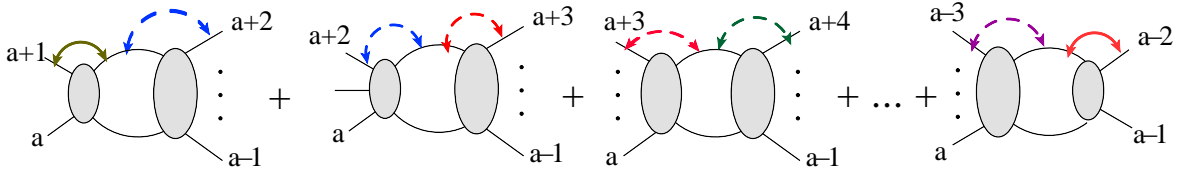}
\caption{A schematic representation of the cancellation of CCP for adjacent MHV amplitude for SYM. Each colored arrow represents a collinear pole that contributes to the bubble coefficient. Pairs of dashed arrows in the same color cancel. Only those represented by the solid arrows one on the two ends remain; they are the only non-trivial contribution to the overall bubble coefficient.}
\label{crust}
\end{center}
\end{figure}

The cancellation of CCP in adjacent channels leads to systematic cancellation in the sum of bubble coefficients, and the sum telescopes. However, there are ``terminal cuts" which contain unique poles that are not cancelled. Below, we demonstrate that these so-called ``terminal poles'' constitute the sole contribution to the overall bubble coefficient. First we focus on the simplest case, namely the two $\Psi$-lines $a,b$ are adjacent. The general case is treated in section.~\ref{NonAdjSect}.

\subsubsection{Adjacent MHV amplitudes}\label{adjsect}

We consider split-helicity MHV amplitudes where the $\Psi$ lines $a,b$ are adjacent, i.e. $b=a-1$.
The systematic cancellation is illustrated in Fig.~\ref{crust}, where the dashed lines indicate pairs of CCP that cancel in the sum.
Note that there are no contributions from the collinear poles where the loop leg is collinear with the $\Psi$-lines, $a$ and $a-1$.
This is because the residues of such poles are zero, as can be seen in eq.~(\ref{factor}) and explicitly checked in eq.~(\ref{Mess}).
One immediately sees that the summation is reduced to the two terminal poles.
These are identified as poles in two-particle cuts with a 4-point tree amplitude on one side (and an $n$-point tree on the other), where the two loop momenta become collinear with the two external legs of the 4-point tree amplitude.
A straightforward evaluation of the contribution of these two terminal poles yields the result for the sum of all bubble coefficients:
\eqa
\nonumber {\cal C}_{2\, \{a,a-1\}} &=& {C}_{2\, \{a,a-1\}}^{(a-3,a-1)}(\l \sim a-2) + {C}_{2\, \{a,a-1\}}^{(a+1,a-1)}(\l \sim a+1)\\
&=&-({ \cal N}-4){\cal A}_n^{\text{tree}} + 0= -\beta_0 {\cal A}_n^{\text{tree}}\,,
\eqae
with $\beta_0=({\cal N}-4)$. Note that ${C}_{2\, \{a,a-1\}}^{(a+1,a-1)}(\l \sim a+1)=0$ is a result of our choice of reference spinor $|\alpha\>=|a\>$ in deriving eq.~(\ref{Mess}).
Were we to make the other choice, $|\alpha\>=|b\>=|a-1\>$, we would instead have $ {C}_{2\, \{a,a-1\}}^{(a-3,a-1)}(\l \sim a-2)=0$ and ${C}_{2\, \{a,a-1\}}^{(a+1,a-1)}(\l \sim a+1)=-\beta_0 {\cal A}_n^{\text{tree}}$.

\begin{figure}[htbp]
\begin{center}
\includegraphics{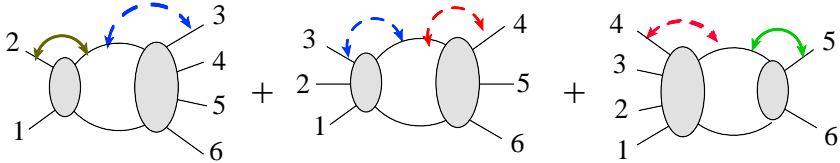}
\caption{A schematic representation of the cancellation of CCP for adjacent six-point MHV amplitude. The dashed lines are common collinear poles, which cancel pairwise.}
\label{6ptexp}
\end{center}
\end{figure}

For example take the six-point MHV amplitude with legs $1$ and $6$ to be negative-helicity lines. The sum of bubble coefficient is given as
\eqa
\nonumber {\cal C}_{2\, \{1,6\}} &=& {C}_{2\, \{1,6\}}^{(2,6)}(\l \sim 2) + {C}_{2\, \{1,6\}}^{(2,6)}(\l \sim 3)+ {C}_{2\, \{1,6\}}^{(3,6)}(\l \sim 3)+ {C}_{2\, \{1,6\}}^{(3,6)}(\l \sim 4)\\
\nonumber&&+ {C}_{2\, \{1,6\}}^{(4,6)}(\l \sim 4)+ {C}_{2\, \{1,6\}}^{(4,6)}(\l \sim 5) \\
&=& {C}_{2\, \{1,6\}}^{(2,6)}(\l \sim 2)  + {C}_{2\, \{1,6\}}^{(4,6)}(\l \sim 5)= -\beta_0 {\cal A}_n^{\text{tree}}\,.
\eqae
We see that there are two pairs of common collinear poles, $\l \sim 3$ and $\l \sim 4$. The pairs cancel each other in the sum and one arrives at the two terminal pole which evaluates to the desired result. The cancellation is illustrated in Fig.~\ref{6ptexp}.

Thus we have demonstrated that one of the terminal poles vanishes and \textit{the sum of bubble coefficients for adjacent MHV amplitude, with arbitrary $n$, is given by a single terminal pole!}
\subsubsection{Non-adjacent MHV amplitudes}\label{NonAdjSect}
The above case with the two $\Psi$-lines $a,b$ being adjacent  is simple because the j-channel poles (see below \eqref{factor}) were absent.
For MHV amplitudes with $a,b$ being non-adjacent, the $j$-channel poles are now non-zero, and all the four collinear poles contribute in eq.~\eqref{Mess}.
The sum of bubble coefficients can be conveniently separated into a summation of the $i$-channel poles, and a summation of the $j$-channel poles.
Cancellation of CCP in both channels again reduces the summation to the terminal poles. For simplicity, we set $a=1$ and $1<b$. We denote the terminal cut in the summation of $i$-channel poles by $i_t,j$, and similarly for the terminal cut of the $j$-channel poles by $i,j_t$.
The contribution of these uncancelled poles are identified as:
\begin{itemize}
  \item $i$-channel:
  $$C^{(i_t,j)}_{2 \ \{1,b\} }(\lambda\sim i_t)\;\;\left.\begin{array}{cc}\;\;{\rm for}\;\; j=n, & i_t=2 \\\;\;{\rm for}\;\;  b\leq j<n, & i_t =1\end{array}\right.\,,$$
  $$C^{(i_t,j)}_{2 \ \{1,b\} }(\lambda\sim i_t +1)\;\;\left.\begin{array}{cc} \;\;{\rm for}\;\; j=b, & i_t=b-2 \\\;\;{\rm for}\;\;   b < j  \leq n, & i_t=b-1\end{array}\right.\,,$$
  \item $j$-channel:
  $$C^{(i,j_t)}_{2 \ \{1,b\} }(\lambda\sim j_t)\;\;\left.\begin{array}{cc}\;\;{\rm for}\;\;  i=b-1, & j_t=b+1 \\ \;\;{\rm for}\;\; 1\leq i<b-1, & j_t =b\end{array}\right.\,,$$
  $$C^{(i,j_t)}_{2 \ \{1,b\} }(\lambda\sim j_t +1)\;\;\left.\begin{array}{cc}\;\;{\rm for}\;\;  i=1, & i_t=n-1 \\ \;\;{\rm for}\;\;  1 < i  \leq n-1, & j_t=n\end{array}\right.\,.$$
\end{itemize}
In identifying the terminal poles, one has to take into account that, when summing over the $i$-channel poles, the value of $j$ affects the possible values that $i$ can take (and vice versa for the summation of $j$-channel poles). For a detailed discussion of the above result, we refer the reader to appendix \ref{NAMHV} where we perform a similar analysis for non-adjacent MHV amplitudes in pure Yang-Mills. As discussed in section~\ref{adjsect}, the collinear poles where the loop momenta becomes collinear with a $\Psi$-line have vanishing residues. In the present context, this refers to $(\lambda\sim 1)$ and $(\lambda\sim b)$. Thus there are only four contributing terms in the sum of bubble coefficients
\eq
{\cal C}_{2\,\{1,b\}}=C^{(2,n)}_{2 \ \{1,b\} }(\lambda\sim 2)+C^{(b-2,b)}_{2 \ \{1,b\} }(\lambda\sim b-1)+C^{(b-1,b+1)}_{2 \ \{1,b\} }(\lambda\sim b+1)+C^{(1,n-1)}_{2 \ \{1,b\} }(\lambda\sim n)\,.
\label{AWTY}
\eqe
Extracting the corresponding expressions from eq.~(\ref{Mess}), one finds that the first and last terms vanish. This is again due to the choice of reference spinor $| \a \> = |a \>$. \fn{If we were to use the other choice, $| \a \> = | b \>$, we would arrive at the result that the second and third terms of eq.~(\ref{AWTY}) vanish. This apparent dependence of a particular double-cut on the reference spinor is illusory: with care, one can cancel the full $|\a\>$-dependence from each individual bubble coefficient.  However, this cancellation comes at the expense of the manifest $a \leftrightarrow b$ symmetry present in the uncancelled form.  This asymmetry causes one term to seemingly vanish while the other gives the full bubble-coefficient.
} Thus the only contributions to the sum of bubble coefficients come from ${C}_{2\;\{1,b\}}^{(b-1,b+1)}(\l \sim b+1)$ and ${C}_{2\;\{1,b\}}^{(b-2,b)}(\l \sim b-1)$, which sum to
\eqa
\nonumber {\cal C}_{2\,\{1,b\}}\hspace{-2mm} &=& \hspace{-2mm} {C}_{2 \ \{1,b\}}^{(b-1,b+1)}(\l \sim b+1) + {C}_{2 \ \{1,b\} }^{(b-2,b)}(\l \sim b-1)\\
&=&\hspace{-2mm} (4-{\cal N}) {\cal A}_n^{\text{tree}}\frac{\<1, b-1\>\<b, b+1\>+\<b-1, b\>\<1, b+1\>}{\<b-1, b+1\>\<1, b\>} = -({ \cal N}-4) {\cal A}_n^{\text{tree}}\,.
\eqae
This agrees with eq.~(\ref{constraint}) with $\beta_0=({ \cal N}-4)$.

In conclusion, for both adjacent and non-adjacent MHV amplitudes in ${\cal N} = 1, 2$ super Yang-Mills theory, the cancellation of CCP in the sum of bubble coefficients implies that \textit{for $n$-point (non-)adjacent MHV amplitudes, only (two) one term in the sum of bubble coefficients gives a non-trivial contribution $\beta_0{\cal A}_n^{\rm{tree}}$.} Thus the on-shell formalism achieves eq.~(\ref{constraint}) in a systematic and simple way.

\subsection{MHV bubble coefficients for pure Yang-Mills}\label{pYM}

The observed structure of cancellations for $\mathcal{N}=1,2$ super Yang-Mills theory is present in pure Yang-Mills as well. However, it is more involved to derive this since the $\mathcal{O}(z^0)$ part of the BCFW-shifted two-particle cut contains higher-order collinear poles. Nevertheless, adjacent channels again share these higher-order CCP, and their contribution to the sum of bubble coefficient also cancels. The cancellation of CCP renders the summation down to the terminal poles, which evaluate to $11/6A_n^{\rm tree}$. We present the detailed derivation of this in appendix \ref{NAMHV}. Here we would like to give a brief discussion on the nature of the terminal poles in pure Yang-Mills theory.

As discussed above, the terminal cuts are those where there is a 4-point tree amplitude on one side of the two-particle cut.
The uncancelled terminal poles can be identified as the poles that arise when the loop momenta become collinear with the pair of external legs of this 4-point tree amplitude.
For pure Yang-Mills, summing over the internal helicity configurations \emph{before} taking the $dz$ and $d{\rm LIPS}$ integrals obscures the nature of the cancellation.

\begin{figure}
\begin{center}
\includegraphics[width=5in]{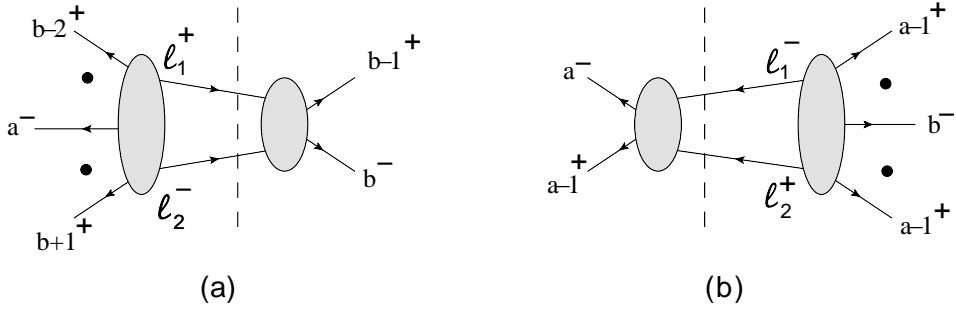}
\caption{The two terminal cuts for a given helicity configuration for the loop legs. For the choice of reference spinor $|\alpha\>=|a\>$ only diagram (a) is non vanishing. If one instead choose  $|\alpha\>=|b\>$, then it is diagram (b) that gives the non-trivial contribution.}
\label{TargetPole}
\end{center}
\end{figure}

Additional structure reveals itself if we \emph{first} evaluate the contributions to the bubble-coefficient for a given set of internal states, aka gluon helicity configuration, and \emph{then} sum over internal states/helicities.
Specifically, these double-forward terminal poles are non-zero only when the internal helicities of the loop legs leaving the $n$-point tree on one side of the cut, match with the helicities of the pair of external lines in the 4-point tree on the other side of the cut (see Fig.~\ref{TargetPole}).
These ``helicity preserving'' double-poles (which will henceforth be called ``double-forward poles'') give the entire bubble coefficient.

Consider the internal helicity configuration ($l_1^+,l_2^-$) as shown in Figs.~\ref{pmmp} and~\ref{TargetPole}.
There are two ``helicity preserving'' terminal-cuts: diagram (a) and (b) in Fig.\ref{TargetPole}.
Choosing the reference spinor $|\a\>=|a\>$, diagram (b) vanishes, and diagram (a) evaluates to $11/6A_n^{\rm tree}$, see eq.~\eqref{Target}.
If one were to make the other choice for the reference spinor, $|\a\>=|b\>$, we would instead have diagram (a) in Fig~\ref{TargetPole} vanishing, and diagram (b) giving $11/6A_n^{\rm tree}$.
In fact, the helicity preserving property of the contributing poles can also be seen for the $\mathcal{N}=1,2$ super Yang-Mills theory, where one simply substitute the $+$ and $-$ helicity in the previous discussions with $\Psi$ and $\Phi$ lines.
This fact was obscured previously as the different internal multiplet configurations were summed to obtain the simple form of the two-particle cut in eq.~(\ref{S}).

Thus we conclude that in the pure Yang-Mills theory \textit{the sum of bubble coefficients is simply given by the contribution of terminal poles where the helicity configuration is preserved, and where the loop momenta become collinear with the pair of external legs within the 4-point amplitude.} For simplicity we will call these double-forward poles, due to the nature of the kinematics. In section \ref{NMHV_Sect} we will show that for split-helicity N$^k$MHV bubble coefficients, these double-forward poles again produce the correct sum for the bubble coefficient, thus indicating complete cancellation among the remaining contributions.  But before indulging in that story let us present a general argument for the cancellation of CCP.

\section{Towards general cancellation of common collinear poles}\label{CCCP}
In the above, we have shown that eq.~(\ref{constraint}) can be largely attributed to the fact that the bubble coefficient for a given cut, secretly shares the same terms with it's four adjacent cuts,  leading to systematic cancellations between them. We have proven this for $n$-point MHV amplitudes in both $\mathcal{N}=1,2$ super Yang-Mills as well as pure Yang-Mills theories (appendix~\ref{NAMHV}). One can also consider adjoint scalars and fermions minimally coupled to gluons. Since at one-loop, we can separate contributions from different spins inside the loop as
\eqa
\nonumber \text{fermions} &~\rightarrow~& \big(~\mathcal{N}=1\;{\rm SYM}~\big)~-~\big(~{\rm YM}~\big)\\
\text{scalar} &~\rightarrow~&\big(~ \mathcal{N}=2\;{\rm SYM} ~\big)~ -  ~\big(~\mathcal{N}=1\;{\rm SYM}~~\big)~ - ~\big(~{\rm fermions}~\big)\,,
\label{SusyDecomp}
\eqae
proof of cancellation of CCP for each of the theories (for MHV scattering) on the RHS of eq.~\eqref{SusyDecomp}, implies that such cancellation occurs for each spin individually.

We would like to show this holds for N$^k$MHV amplitudes.  Unfortunately, for N$^k$MHV amplitudes, multi-particle poles of tree amplitudes on either side of the cut contribute to the bubble coefficient, and the analysis becomes more complicated. However, we believe that the cancellation between CCP persists for arbitrary helicity configuration. As an indication, we demonstrate that the residues of CCP for adjacent cut always have the same form and opposite signs, for any helicity configuration.

Collinear limits of tree level amplitudes in Yang-Mills theory, with $k_a=zk_P, \;k_b=(1-z)k_P$, factorize as
\eq
A_n^{\text{tree}}\left(...,a^{\lambda_a},b^{\lambda_b},...\right)\rightarrow \sum_{\lambda=\pm}{\rm  Split}_{-\lambda}^{\text{tree}}\left(z,a^{\lambda_a},b^{\lambda_b} \right) A_{n-1}^{\text{tree}}\left(...,P^{\lambda},...\right)
\eqe
where the factor ${\rm  Split}_{-\lambda}^{\text{tree}}\left(z,a^{\lambda_a},b^{\lambda_b} \right)$ is the gluon splitting amplitude.  Its form for various helicity configurations are given by~\cite{ParkeTaylor, Dixon:1996wi}:
\eqa
{\rm  Split}_{-}^{\text{tree}}\left(a^{-},b^{-} \right)&=&0\\
\nonumber {\rm  Split}_{-}^{\text{tree}}\left(a^{+},b^{+} \right)&=&\frac{1}{\sqrt{z(1-z)}\langle ab\rangle}\\
\nonumber {\rm  Split}_{+}^{\text{tree}}\left(a^{+},b^{-} \right)&=&\frac{(1-z)^2}{\sqrt{z(1-z)}\langle ab\rangle}\\
\nonumber {\rm  Split}_{-}^{\text{tree}}\left(a^{+},b^{-} \right)&=&-\frac{z^2}{\sqrt{z(1-z)}[ab]}\,.
\eqae
Without loss of generality, we focus on the common collinear pole, depicted in Fig.~\ref{AdjacentCuts}, in adjacent cuts $(1...i-1|i...n)$ and $(1...i|i+1...n)$.
In other words, we study the collinear region with $ l^{(i)}_1=\tau^{(i)} k_i$ and $l^{(i+1)}_1=\tau^{(i+1)} k_i$.\footnote{Strictly speaking, the condition $\langle l_1i\rangle=0$ only requires $\lambda_{l_1}\sim\lambda_i$. However since the $d{\rm LIPS}$ integration contour is along $\tilde\lambda=\overline{\lambda}$, the condition is equivalent to $l_1\sim k_i$.} The two integrands become
\eqa
&&\hspace{-16mm}{\rm Cut}_{(1..i-1|i..n)}|_{\langle l_1i\rangle}\label{cut1}\\
 \nn&=&A_{i+1}\left(1,...,i-1,\tau^{(i)}i,l^{(i)}_2\right) \sum_{\lambda=\pm}{\rm  Split}_{-\lambda}^{\text{tree}}A_{n-i+2}\left(-l^{(i)}_2,(1-\tau^{(i)})i,i+1,...,n\right) \ ,
\eqae
for cut $(1...i-1|i...n)$, and
\eqa
&&\hspace{-11mm} {\rm Cut}_{(1..i|i+1...n)}|_{\langle l_1i\rangle}\label{cut2}\\
\nonumber&=&\sum_{\lambda=\pm}{\rm  Split}_{-\lambda}^{\text{tree}}A_{i+1}\left(1,...,(1+\tau^{(i+1)})i,l^{(i+1)}_2\right) A_{n-i+2}\left(-l^{(i+1)}_2,-\tau^{(i+1)}i,i+1,...,n\right) \  ,
\eqae
for cut $(1...i|i+1...n)$. The parameter $\tau^{(i)}$ can be fixed by the on-shell condition on $l^{(i)}_2$ since in the cut $(1...i-1|i...n)$, $l^{(i)}_2 = P_{i-1}+\tau^{(i)}k_i$. Similar constraints from the cut $(1...i|i+1...n)$ fix $\tau^{(i+1)}$. This leads to
\eqa
\nonumber &&\tau^{(i)}=\frac{P^2_{i-1}}{2k_i\cdot P_{i-1}}=\tau^{(i+1)}+1\\
\nonumber &\rightarrow& l^{(i)}_2 = P_{i-1}+ \tau^{(i)} k_{i}=P_{i}+ \tau^{(i+1)} k_{i}=l^{(i+1)}_2\,.
\eqae
Substituting these results back into eq.~(\ref{cut1}) and eq.~(\ref{cut2}), we see that the product of tree amplitudes are identical at their common collinear pole. Furthermore, identifying the kinematic variables in the splitting amplitudes for each cut as:
\eqa
\nonumber (1...i-1|i...n):&& k_a=k_i,\;k_b=-\tau^{(i)}k_i, \;z=\frac{1}{1-\tau^{(i)}}\\
\nonumber (1...i|i+1...n):&&k'_a=k_i,\;k'_b=\tau^{(i+1)}k_i, \;z'=\frac{1}{1+\tau^{(i+1)}}=\frac{1}{\tau^{(i)}}\,
\eqae
we see that the splitting amplitudes for the two cuts are identical with a relative minus sign.\footnote{For consistency, we take the positive branch of the square root.}

The above analysis confirms that eq.~(\ref{cut1}) and eq.~(\ref{cut2}) are indeed identical up to a minus sign. Thus the residue of the \textit{entire two-particle cut} on the common collinear poles, are identical and with opposite sign. This, however, does not directly lead to a proof of cancellation of CCP for bubble coefficients. This is because to extract the bubble coefficient, the two-particle cut must be translated into a total derivative, in order for one to use holomorphic anomaly generated by the collinear poles to isolate the $d{\rm LIPS}$ integral. It is not guaranteed that after translating the two cuts into a total derivative form, the residues on the CCP are still equal and opposite.

\section{N$^{k}$MHV bubble coefficients}\label{NMHV_Sect}

The cancellation of CCP, even if it holds for generic helicity configurations, is clearly not sufficient for simplifying the sum of bubble coefficients for N$^k$MHV amplitudes. The complications arise from the presence of multi-particle poles of the tree amplitudes in the two-particle cut. It is conceivable that there exists a similar cancellation of common multi-particle poles, since a trivial example would be the cancellation of CCP, considering the fact that  collinear poles are secretly multi-particle poles via momentum conservation.
Here we instead ask a more direct question: does the contribution of the double forward poles for the bubble coefficient, directly gives $11/6A^\text{tree}$ for $n$-point N$^k$MHV amplitude.

To facilitate our analysis, we will use the CSW representation~\cite{csw, Risager:2005vk, efk} for the split helicity NMHV tree amplitude $(---+\cdot\cdot\cdot+)$. We will show that at the double forward poles, the contribution from each individual CSW diagram evaluates to $11/6$ times the original CSW diagram whose loop momenta is replaced by the corresponding external lines. Summing the different diagrams one simply recovers $11/6$ times the CSW representation of the tree amplitude. Using induction, we prove that this is still true for all $n$-point split-helicity N$^k$MHV amplitudes.

\subsection{Double forward poles in terminal cuts of $A^{1-{\rm loop}}_n(---+...+)$}
The split helicity configuration for NMHV amplitudes is the simplest to analyze. The CSW form for $n$-particle NMHV scattering, with adjacent negative-helicity gluons is given by the following $2(n-3)$ terms~\cite{csw}:
\begin{figure}
\begin{center}
\includegraphics[width=4.5in]{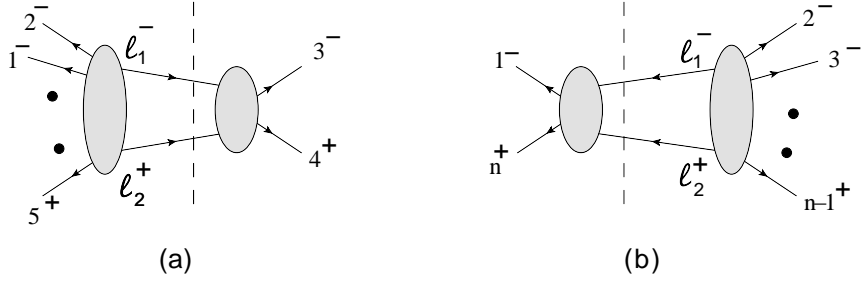}
\caption{The terminal cuts of the split helicity NMHV amplitude that contain the two helicity-preserving double-forward poles.}
\label{CSWTarget}
\end{center}
\end{figure}
\eqa
A(1^-,2^-,3^-,4^+,...,n^+)
&=& \sum_{i=4}^n \frac{\<1,2\>^3}{\< P_{3,i}, i+1 \> \cdot\cdot\cdot\< P, 2 \>} \frac{1}{P^2_{3,i}} \frac{\< P_{3,i}, 3 \>^3}{\<3, 4 \>\cdot\cdot\cdot \< i, P_{3,i} \>} \nonumber\\
&+& \sum_{i=4}^n \frac{\<1,P_{2,i-1}\>^3}{\< P_{2,i-1}, i \> \cdot\cdot\cdot \< n, 1 \>} \frac{1}{P^2_{2,i-1}} \frac{\< 2, 3 \>^3}{\< P_{2,i-1}, 2 \>\cdot\cdot\cdot \< i-1, P_{2,i-1} \>} \nonumber\\
\label{CSWTree}
\eqae
where $|P_{i,j}\> \equiv P_{i,j} | \tilde{\eta}]$, and $\tilde{\eta}$ is an auxiliary spinor.

\begin{figure}
\begin{center}
\includegraphics[width=4.5in]{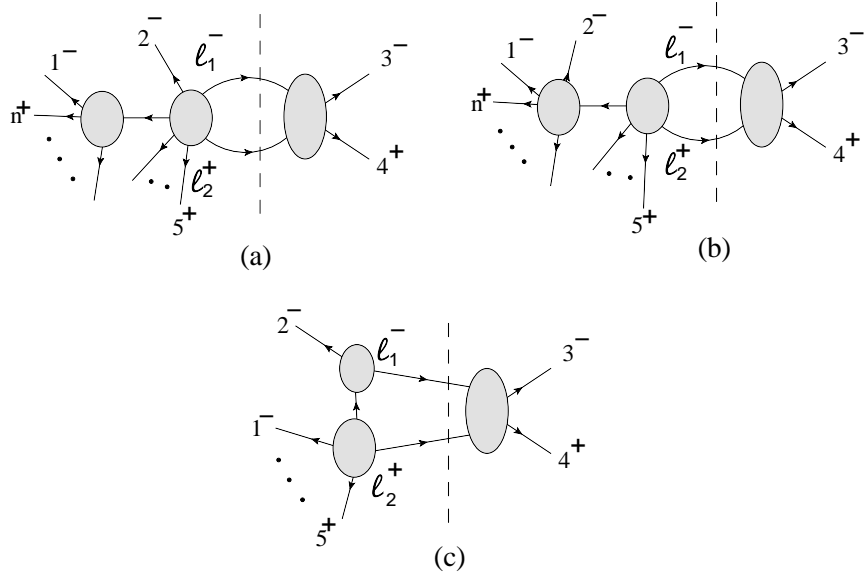}
\caption{Representing the tree amplitude in the terminal cuts with CSW expansion. Note that the loop legs are attached to the same MHV vertex for diagram (a) and (b). The $d{\rm LIPS}$ integration for these diagrams are exactly the same as that computed for adjacent MHV bubbles.}
\label{CSWCh}
\end{center}
\end{figure}

For helicity configuration $(l_1^-,l_2^+)$, the terminal cuts are shown in Fig.~\ref{CSWTarget}. We first focus on cut (a), which is given by $A_4(\hat{l}^+_1, 3^-,4^+,\hat{l}_2^-)$ $\times$$ A_n(\hat{l}_2^+, 5^+,...,n^+,...,1^-,2^-,\hat{l}_1^-)$.  We use the CSW expansion on the $n$-particle NMHV sub-amplitude as indicated in Fig.~\ref{CSWCh}. Notice for diagram (a) and (b) of Fig.~\ref{CSWCh}, the loop legs are on the same MHV vertex and hence the CSW propagator $1/P^2$ does not depend on $z$. This implies that from the point of view of extracting the pole at $z\rightarrow\infty$ and performing the $d{\rm LIPS}$ integration, only the MHV vertex on which the loop legs are attached are relevant. \textit{Hence the evaluation of the double forward poles is simply evaluating the contribution of that for an MHV amplitude with one on-shell leg identified with CSW propagator leg, multiplied by the remaining CSW vertices which behaves as spectators to the $d{\rm LIPS}$ integration.} This realization makes the computation trivial, as we know the double forward pole contributes 11/6 times the tree amplitude. This implies that here the result would simply be 11/6 times the corresponding CSW tree diagram. One can thus straightforwardly obtain:
\eqa
[\; \text{diag}(a)+\text{diag}(b)\;]\bigg|_{df}&=&\frac{11}{6}\bigg(\sum_{i=4}^n \frac{\<1,2\>^3}{\< P_{3,i}, i+1 \> \cdot\cdot\cdot\< P, 2 \>} \frac{1}{P^2_{3,i}} \frac{\< P_{3,i}, 3 \>^3}{\<3, 4 \>\cdot\cdot\cdot \< i, P_{3,i} \>} \nonumber\\
\nonumber&+& \sum_{i=5}^n \frac{\<1,P_{2,i-1}\>^3}{\< P_{2,i-1}, i \> \cdot\cdot\cdot \< n, 1 \>} \frac{1}{P^2_{2,i-1}} \frac{\< 2, 3 \>^3}{\< P_{2,i-1}, 2 \>\cdot\cdot\cdot \< i-1, P_{2,i-1} \>} \bigg)\,,\\
\eqae
where $|_{df}$ indicates the contribution from the double forward pole. There is another type of contribution, as shown in diagram (c) of Fig.~\ref{CSWCh}, where the CSW propagator depends on $z$ and we need a careful analysis as following.

Denoting $|\hat{P}\> = \hat{P} | \tilde{\eta}]$, which accounts for the $z$-dependence of the CSW propagator, the cut is given by:
\eqa
\text{diag}(c)= \frac{\<1, \hat{P} \>^3}{\<\hat{P}, l_2 \> \<  \hat{l}_2, 5 \> ... \< n, 1 \>} \frac{1}{\hat{P}^2} \frac{\< 2,  \hat{l}_1\>^3}{\<\hat{P}, 2 \> \<  \hat{l}_1, \hat{P}\>}
\times
\frac{\<3,  \hat{l}_2\>^4}{\<3, 4 \> \< 4, l_2 \> \<  l_2,  \hat{l}_1 \> \<  \hat{l}_1, 3 \>}
\label{hard1}
\eqae
where $\hat{P} = p_2 +  \hat{l}_1$.  We call this the ``hard'' term.  The other $2n - 5$ terms we call the ``easy'' terms. For simplicity, we first strip off the corresponding tree factor, i.e. the $i=4$ term in the second line of eq.~(\ref{CSWTree}):
\eqa
T_{2,3} = \frac{\<1, P_{2,3} \>^3}{\<P_{2,3}, 4 \> \< 4, 5 \> ... \< n, 1 \>} \frac{1}{P_{2,3}^2} \frac{\< 2, 3\>^3}{\<P_{2,3}, 2 \> \< 3, P_{2,3}\>}
\label{hardTree}
\eqae
Stripping off $T_{2,3}$ from eq.~(\ref{hard1}) yields:
\eqa
\text{diag}(c) =T_{2,3} \bigg( \frac{\<1, \hat{P} \>^3\<P_{2,3}, 4 \> \< 4, 5 \> }{\<1, P_{2,3} \>^3\<\hat{P}, l_2 \> \<  l_2, 5 \>} \frac{P_{2,3}^2}{\hat{P}^2} \frac{\< 2,  \hat{l}_1\>^3}{\<\hat{P}, 2 \> \<  \hat{l}_1, \hat{P}\>}\frac{\<P_{2,3}, 2 \> \< 3, P_{2,3}\>\<3,  l_2\>^4}{\< 2, 3\>^3\<3, 4 \> \< 4, l_2 \> \<  l_2,  \hat{l}_1 \> \<  \hat{l}_1, 3 \>}\bigg)
\label{hard2}
\eqae
We would like to put this into a form where we can readily take the large-$z$ pole followed by the $d{\rm LIPS}$ integral about the double forward pole.  To simplify the analysis, we work-out the explicit form of the spinor-inner products:
\eqa
| \hat{P}\> = (p_2 + \hat{l_1}) | \tilde{\eta}] \Rightarrow
& &\hspace{-4mm} \bigg\{
\< \hat{l}_1, \hat{P}\> = \< \hat{l}_1, 2 \> [2, \tilde{\eta}], \qquad \< l_2, \hat{P}\> = \< l_2, 2\> [2, \tilde{\eta}] + \< l_2, \hat{l}_1 \> [l_1, \tilde{\eta}], \nonumber\\
& & \ \ \<2,\hat{P} \> = \<2, \hat{l}_1 \> [l_1, \tilde{\eta}], \qquad \< 1, \hat{P}\> = \< 1, \hat{l}_{1} \> [l_1, \tilde{\eta}] + \< 1, 2 \> [2, \tilde{\eta}]
\bigg\} \nonumber\\
\nonumber| P_{2,3}\> = (p_2 + p_3) | \tilde{\eta}] \Rightarrow
& &\hspace{-4mm} \bigg\{
\< 3, P_{2,3}\> = \< 3, 2 \> [2, \tilde{\eta}], \qquad \<l_2, P_{2,3}\> = \< l_2, 2 \> [2, \tilde{\eta}] + \<l_2,3 \> [3, \tilde{\eta}], \nonumber\\
\nonumber& & \ \<  2, P_{2,3} \> = \< 2,3 \> [3, \tilde{\eta}], \qquad \<1, P_{2,3}\> = \< 1, 3 \> [3, \tilde{\eta}] + \< 1, 2 \> [2, \tilde{\eta}]
\bigg\}\,.\,\,\\
\label{hardCSW}
\eqae
Applying eq.~(\ref{hardCSW}) to eq.~(\ref{hard2}), cancelling all common factors, and setting $|\tilde{\eta}] = | 2]$,\fn{This ordering of steps is important: this cancels an apparent factor of $[2, \tilde{\eta}]$ in the denominator of eq.~\eqref{hard2}.} we find:
\eqa
\nonumber \text{diag}(c) &=&T_{2,3}\frac{\<1\hat{l}_1\>^3\<4,5\>\<3,l_2\>^4}{\langle13\rangle^3\<\hat{l}_1l_2\>^2\< l_2,5\>\<4,l_2\>\<\hat{l}_1,3\>}\\
&=& T_{2, 3} \bigg[
\bigg(\frac{\<3, 4\>}{\< l_2,4 \>} - \frac{\< 3,5 \>}{\<l_2, 5 \>} \bigg)
\frac{\<\hat{l}_1, l_2 \>}{\<\hat{l}_1, 3 \>}
\bigg( -1 + \frac{\< 1, l_2\> \<\hat{l}_1,3 \>}{\< 1, 3 \> \< \hat{l}_1, l_2 \>} \bigg)^3
\bigg] \ .
\eqae
Expanding around $z\rightarrow\infty$, one finds
\eqa
\nonumber \text{diag}(c)|_{z\rightarrow\infty} &=& T_{2, 3} \bigg[
\bigg(\frac{\<3, 4\>}{\< l_2,4 \>} - \frac{\< 3,5 \>}{\<l_2, 5 \>} \bigg)
\bigg(3 \frac{\< 1, l_2\> }{\< 1, 3 \>} -3\frac{\< 1, l_2\>^2 \<l_1,3 \>}{\< 1, 3 \>^2 \< l_1, l_2 \>}+\frac{\< 1, l_2\>^3 \<l_1,3 \>^2}{\< 1, 3 \>^3 \< l_1, l_2 \>^2}\bigg)
\bigg] \ .
\eqae
The double forward pole corresponds to the $\< l_2,4 \>$ pole. A straightforward evaluation of the residue gives:
\eqa
\nonumber \text{diag}(c)|_{df} &=& T_{2, 3} \int d{\rm LIPS}\;\frac{\<3, 4\>}{\< l_2,4 \>}
\bigg(3 \frac{\< 1, l_2\> }{\< 1, 3 \>} -3\frac{\< 1, l_2\>^2 \<l_1,3 \>}{\< 1, 3 \>^2 \< l_1, l_2 \>}+\frac{\< 1, l_2\>^3 \<l_1,3 \>^2}{\< 1, 3 \>^3 \< l_1, l_2 \>^2}\bigg)\\
\nonumber &=& \left(3\times1-3\times\frac{1}{2}+1\times\frac{1}{3}\right)T_{2, 3}=\frac{11}{6}T_{2, 3}\,.
\eqae

In summary, we find that the double forward pole of the terminal cuts sum to give:
\eqa
[\;\text{diag}(a)\;+&\text{diag}(b)&+\;\text{diag}(c)\;]\bigg|_{df}=\frac{11}{6}\bigg(\sum_{i=4}^n \frac{\<1,2\>^3}{\< P_{3,i}, i+1 \> \cdot\cdot\cdot\< P, 2 \>} \frac{1}{P^2_{3,i}} \frac{\< P_{3,i}, 3 \>^3}{\<3, 4 \>\cdot\cdot\cdot \< i, P_{3,i} \>} \bigg)\nonumber\\
&+&\frac{11}{6}\bigg( \sum_{i=5}^n \frac{\<1,P_{2,i-1}\>^3}{\< P_{2,i-1}, i \> \cdot\cdot\cdot \< n, 1 \>} \frac{1}{P^2_{2,i-1}} \frac{\< 2, 3 \>^3}{\< P_{2,i-1}, 2 \>\cdot\cdot\cdot \< i-1, P_{2,i-1} \>} \bigg)\nonumber\\
&+&\frac{11}{6}\bigg(\frac{\<1, P_{2,3} \>^3}{\<P_{2,3}, 4 \> \< 4, 5 \> ... \< n, 1 \>} \frac{1}{P_{2,3}^2} \frac{\< 2, 3\>^3}{\<P_{2,3}, 2 \> \< 3, P_{2,3}\>} \bigg)\nonumber\\
\nonumber&=&\frac{11}{6}A(1^-,2^-,3^-,4^+,...,n^+)\\
\label{Proof}
\eqae
Thus indeed the double forward limit gives the expected proportionality factor from the sum of bubble coefficients, suggesting a complete cancellation among the contributions from all the other channels.

We see that using the CSW representation for NMHV tree amplitudes reveals the following structure: the double forward poles in the terminal cut contributes a factor 11/6 for \textit{each} CSW tree diagram. Note that all but one term, diagram (c), goes through trivially as the $d{\rm LIPS}$ only sees only one MHV vertex in the NMHV tree amplitude.
There is a more straightforward way of understanding the factor of 11/6 in diagram (c), which goes as follows.

\begin{figure}
\begin{center}
\includegraphics[width=4.5in]{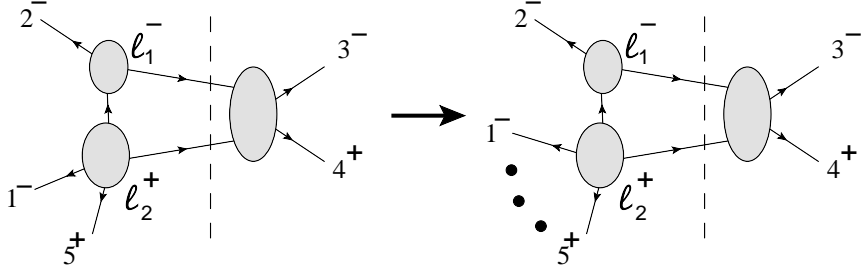}
\caption{Diagram (c) in Fig.~\ref{CSWCh} for the five-point amplitude. Going from five-point to arbitrary $n$-point simply corresponds to adding additional plus helicity legs on the bottom MHV vertex. Since this modification affects neither extraction of constant term for $z\rightarrow\infty$ nor evaluation of the $d{\rm LIPS}$ integral, the $11/6$ factor obtained at five-points holds for arbitrary $n$.}
\label{diagramC}
\end{center}
\end{figure}

Consider the same calculation for the split-helicity NMHV five-point amplitude. Since this is secretly a five-point $\overline{\rm MHV}$ amplitude we know that the forward poles for a given internal helicity configuration evaluate to $11/6\times$ $\overline{\rm MHV}$, as proven previously. Now consider the same evaluation using the CSW representation. Since diagrams (a) and (b) automatically yield 11/6 times the corresponding CSW tree diagram, diagram (c) must give 11/6 times its corresponding CSW tree diagram, $T_{2,3}$.
For higher point NMHV amplitudes, diagram (c) is modified by additional plus helicity legs on one of the MHV vertices, as indicated in Fig.~\ref{diagramC}. From the point of view of expanding around the pole at $z\rightarrow\infty$ and the $d{\rm LIPS}$ integral, these additional plus helicity legs are simply spectators and do not participate. Thus the evaluation of diagram (c) ``must" be 11/6 $T_{2,3}$ as explicitly shown above.  Note that this way of understanding the result of the double forward poles allows us to generalize to N$^k$MHV.

\subsection{Recursive generalization to N$^{k}$MHV bubble coefficients}\label{NNMHV_Sect}

We are now ready to give an inductive proof that the residue at the double-forward poles gives the entire bubble coefficient, $11/3 \times A_n^{\rm tree}$, for general split-helicity N$^k$MHV amplitudes. The proof is as follows:
\begin{itemize}
  \item Using the CSW representation of N$^k$MHV amplitude, the diagrams that appear inside the terminal cut can be categorized by the number of $z$ dependence CSW propagators. For a given $k$ there will be at most $k$ propagators that have non-trivial $z$ dependence. Diagrams that have $p<k$, $z$-dependent CSW propagators will be diagrams that have already appeared in the analysis for N$^p$MHV amplitudes, hence are known to give 11/6 times the corresponding CSW tree diagram.
  \item  There will be a unique diagram that has $k$, $z$-dependent, CSW propagators. To evaluate this diagram, we note that the $k+4$-point split-helicity N$^{k}$MHV amplitude is the same as a $k+4$-point adjacent $\overline{\rm MHV}$ amplitude, for which we know that the forward limit poles gives 11/$6A_n^{\text{tree}}$. In the CSW representation, since all other diagrams already evaluate to 11/6 times the corresponding tree diagram, as discussed in the previous step, this final diagram must as well.
  \item For arbitrary $n$, one simply adds additional positive-helicity legs to MHV vertices.  These extra states do not participate in the expansion around the pole at $z\rightarrow\infty$ or in the $d{\rm LIPS}$ integral. The modification only appears as an overall factor, thus proves that for general $n$ this last diagram also evaluates to 11/6 times the corresponding CSW tree diagram.
  \item Summing all the CSW diagrams in the double-cut, we obtain $11/6 \times A^{\text{tree}}_{n}$ for the forward pole contribution to the bubble coefficient from the terminal cut.
  \item The other internal helicity configuration evaluates in the same way, on the other helicity-preserving double-forward terminal pole, yielding $\cc_2 = 11/3 \times A_n^{\rm tree}$
\end{itemize}
\begin{figure}[h]
\begin{center}
\includegraphics[width=4in]{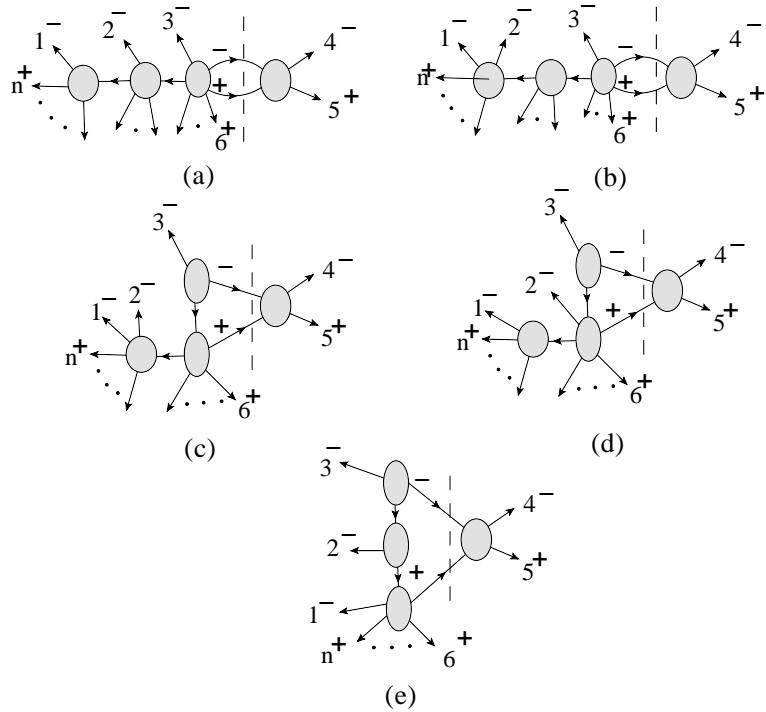}
\caption{The CSW representation of the terminal cut for N$^2$MHV amplitude. The evaluation of diagrams (a) and (b) are identical to that of adjacent MHV amplitudes, while the evaluation of diagrams (c) and (d) are identical to split helicity NMHV amplitude. }
\label{N2MHV}
\end{center}
\end{figure}

We use the N$^2$MHV amplitude to illustrate the above steps. The CSW representation for the N$^2$MHV terminal cut is given in Fig.~\ref{N2MHV}. Diagrams (a) and (b) have no $z$-dependent CSW propagators, and hence from the point of view of extracting the constant piece at $z\rightarrow\infty$ and integrating over $d{\rm LIPS}$, the two left most MHV vertex are just spectators and the evaluation is on the right most MHV vertex on the left hand side of the cut. Thus evaluation of diagrams (a) and (b), is identical to evaluation of adjacent MHV amplitudes. For diagrams (c) and (d), there is one $z$-dependent CSW propagator. The MHV vertex where $1^-$ sits is again a spectator and the evaluation is identical to that of diagram (c) in Fig.~\ref{CSWCh} for the NMHV amplitude, and hence evaluates to 11/6 times the corresponding CSW tree diagram. Finally, for the unique diagram (e), we use the argument that for $n=6$, this is simply the $\overline{\rm MHV}$ amplitude, from which we deduce that this term must also evaluate to 11/6 times the corresponding tree diagram. This result will not be modified for $n>6$, and hence completes the proof.

\section{Conclusion and future directions }
In this paper, we study the proportionality between the sum of bubble coefficients and the tree amplitude, which is required for renormalizability. For theories where Feynman diagram analysis is tractable, such as scalar theory and pure scalar amplitudes of Yukawa theory, we find that the bubble coefficient only receives contributions from a small class of one-loop diagrams. The contribution of each diagram is proportional to a tree-diagram, and hence summing over all one-loop diagrams that give non-trivial contributions, is equivalent to summing over all tree-diagrams. This trivially leads to the proportionality condition. For some amplitudes of the Yukawa theory, the large-$z$ behavior of the two-particle cut becomes intractable via Feynman diagrams, and we instead use helicity amplitudes. By requiring the sum to be proportional to tree amplitude, we find the known renormalization conditions derived from power counting analysis.

For (super)Yang-Mills theory, we show that the bubble coefficient for MHV amplitudes can be organized in terms of their origin as collinear poles, which are responsible for the nontrivial contribution to the $d{\rm LIPS}$ integration, in the two-particle cuts. This representation reveals the existence of systematic cancellation in the sum of bubble coefficient. In particular, the residues of common collinear poles (CCP) cancels, and the sum telescopes down to unique terminal poles. These are poles that arise from cuts that have at least a 4-point tree amplitude on one side of the cut, and the helicity configuration of the internal legs must match that of the two external legs on the four-point tree amplitude, as shown in Fig.~\ref{TargedIn}. We conjecture that these double forward poles are the only non-trivial contribution to the sum of bubble coefficients for any helicity configuration. As further evidence, we explicitly proved that for split helicity $n$-point N$^k$MHV amplitudes, the contribution of each terminal pole indeed give 11/6 times the tree amplitude.

For more generic external helicity configurations, it will be interesting to see how the contributions from the multi-particle poles cancel with each other. 
An even more interesting example would be gravity. It is well-known that pure gravity is one-loop finite~\cite{'tHooft:1974bx}. The bubble coefficient is non-vanishing for generic two-particle cuts, and hence massive cancellation must occur. The lack of color ordering for gravity amplitudes indicate the pole structure that gives rise to the non-trivial contributions for the $d{\rm LIPS}$ integral is more complicated than Yang-Mills: presumably new cancellation mechanisms are required even for MHV amplitudes.

We have demonstrated that the UV divergence of the one-loop gauge theory amplitude is completely captured by the residues of a set of unique collinear poles, i.e. it is controlled by a residue at finite loop momentum value. If the same holds for gravity, then through KLT relations~\cite{Kawai:1985xq} the residue of gravity is intimately tied to gauge theories, and it will be interesting to see how the relationship allows cancellation among terminal residues, leading to the known finiteness result for gravity and it's relationship to BCJ duality~\cite{Bern:2010ue,Bern:2010yg}. Note that the study of tensor bubbles has previously revealed improved UV behavior for gravity amplitudes compared to naive power counting from Einstein-Hilbert action~\cite{Bern:2007xj}. Even though it is well known that gravity is finite at one-loop, a careful analysis of how finiteness is achieved for generic amplitudes may shed light on additional structure, as we have successfully achieved for (super) Yang-Mills amplitudes.

\section*{Acknowledgements}
We thank Nima Arkani-Hamed, Simon Caron-Huot and Michael Kiermaier  for many enlightening discussions. We are especially grateful to Zvi Bern, Lance Dixon and Henriette Elvang for careful reading of our draft and giving us helpful suggestions. YH is grateful for the Institute of Advanced Study for invitation as visiting member. YH would also like to thank Isaac Newton Institute for Mathematical Sciences for organizing the workshop ``Recent Advances in Scattering Amplitudes", during which part of this work was completed. YH was supported by the US Department of Energy under contract
DE--FG03--91ER40662. DAM is supported NSF GRFP grants DGE-1148900 and PHY-0756966. CP is supported by NSF Grant PHY-0953232, and in part by the DOE Grant DE-FG02-95ER 40899.

\appendix

\section{$d{\rm LIPS}$ integrals, via the holomorphic anomaly in 4-dimensions}\label{Holomorphic}

Following~\cite{SimplestQFT,ehp,csw}, we can calculate integrals of the following form,
\begin{eqnarray}
\oint_{\tilde{\l} = \bar{\l}} P^2 \frac{\<\l, d\l \> [\tilde{\l}, d\tilde{\l}]}{\< \l | P | \tilde{\l}]^2} \frac{\prod_{i = 1}^{n} [a_i, \tilde{\l}]}{\< \l | P | \tilde{\l}]^n} g(\l) \ , {\rm where} \ g(\l) = \frac{\prod_{j = 1}^{m} \< b_j, \l \>}{\prod_{k = 1}^{m} \< c_k, \l \>}
\label{dLIPSgeneric}
\end{eqnarray}
where the integral over phase-space ($d{\rm LIPS}$ integral) is really a contour integral over two complex numbers. Two cases are important here: $n = 0$ for scalar- and Yukawa-theory, and $n = 2$ for gauge theories. Note:
\eqa
P^2 \frac{[\tilde{\l}, d\tilde{\l}]}{\< \l | P | \tilde{\l}]^2} =
-d\tilde{\l}^{\dot{\a}}\frac{\partial}{\partial \tilde{\l}^{\dot{\a}}} \bigg( \frac{[\tilde{\l}, \tilde{\eta}] P^2 }{\< \l | P | \tilde{\l}] \< \l | P | \tilde{\eta}]} \bigg) =
-d\tilde{\l}^{\dot{\a}} \frac{\partial}{\partial \tilde{\l}^{\dot{\a}}} \bigg( \frac{[\tilde{\l}| P | \a \>}{\< \l | P | \tilde{\l}] \< \l, \a \>} \bigg) \,.
\label{derivative}
\eqae
where we have introduced reference spinors $| \tilde{\eta}] = P |\a\>$ in order to express the $d{\rm LIPS}$ integration measure as a total derivative.  We further note that integrands of the form \eqref{dLIPSgeneric} can be reduced to this basic measure through repeated differentiation.  Concretely, for $n = 2$:
\eqa
\frac{[I, \tilde{\l}] [J, \tilde{\l}]}{\< \l | P | \tilde{\l}]^4} &=&\frac{1}{6} \tilde{I}^{\dot{\g}} \tilde{J}^{\dot{\b}} \frac{\partial^2}{\partial(\< \l | P)^{\dot{\b}} \partial(\< \l | P)^{\dot{\g}}} \bigg\{ \frac{1}{\< \l | P | \tilde{\l}]^2} \bigg\} \,.
\label{MHVform}
\eqae
For the case of MHV bubble integrands, the only reference spinors are of the form $| I ] = P|i\>$.  Combining \eqref{derivative} and \eqref{MHVform}, and interchanging the order of differentiation, one can re-write the ``$n = 2$'' integrand as a total derivative:
\eqa
\label{LambdaAlpha}
& & \oint_{\tilde{\l} = \bar{\l}} P^2 \frac{\<\l, d\l \> [\tilde{\l}, d\tilde{\l}]}{\< \l | P | \tilde{\l}]^4} \< i | P | \tilde{\l}] \< j | P | \tilde{\l}] g(\l) \nonumber\\
& &
= \frac{1}{6} \oint_{\tilde{\l} = \bar{\l}} \<\l, d\l \>
\bigg(
- d\tilde{\l}^{\dot{\g}} \frac{\partial}{\partial \tilde{\l}^{\dot{\g}}} \bigg[ \frac{[\tilde{\l}| P | \a \> \ g(\l)}{\< \l | P |\tilde{\l}] \< \l , \a \>} \frac{1}{3} \bigg\{
   \frac{\< i | P | \tilde{\l}] \< j | P | \tilde{\l}]}{\< \l | P | \tilde{\l}]^2}
+ \frac{\< i, \a \> \< j, \a \>}{ \< \l, \a \>^2 }
\\
& &
\ \ \ \ \ \ \ \ \ \ \ \ \ \ \ \ \ \ \ \ \ \ \ \ \ \ \ \ \ \ \ \ \ \ \ \ \ \ \ \ \ \ \ \ \ \ \ \ \ \ \ \ \ \
+ \frac{1}{2} \frac{\< i | P | \tilde{\l}] \< j, \a \> + \< i, \a \> \< j | P | \tilde{\l}]}{  \< \l | P | \tilde{\l}] \< \l, \a \> } \bigg\} \bigg] \bigg)\,.
 \nonumber
 \eqae
The last form for the integrand, re-written as a total derivative, vanishes at all points save when it hits a simple pole.  This is because along the integration contour $\tilde{\l} = \bar{\l}$, one has~\cite{csw},
\eq
-d\tilde{\l}^{\dot{\a}}\frac{\partial}{\partial \tilde{\l}^{\dot{\a}}}\frac{1}{\langle\l,\xi\rangle}=-2\pi \bar{\delta}\left(\langle\l,\xi\rangle\right)\,,
\label{delta}
\eqe
\begin{equation}
    \int \<\l d\l\> \bar{\del}(\<\l,\xi\>) B(\l)=iB(\xi)\,.
\end{equation}
Thus the $d{\rm LIPS}$ integral is localized to the poles $1/\langle\l,\xi\rangle$ of the integrand.\footnote{Note that $1/\< \l | P | \tilde{\l}]$ is not a simple pole on the contour $\tilde{\lambda}=\overline{\lambda}$.}
Each term in the integrand \eqref{LambdaAlpha} has potential collinear divergences coming from the spinor brackets in the denominator of $g(\l)$ and that of the reference spinor $\< \l , \a \> \rightarrow 0$.

Through \eqref{derivative} and \eqref{delta}, we see that the simple bubble integrals in scalar QFT in section~\ref{scalarQFT} simply evaluate to $\int d{\rm LIPS} (1) = -2 \pi i$.
For MHV bubble integrands, such as eq.~\eqref{factor}, there is always a choice of the reference spinor, such as $| \a \> = | a \>$, which eliminates the unphysical $1/\langle \l,\alpha\rangle$ pole.

\section{A proof of \ref{treebubb0} for $n>4$ case}\label{largenscalarpf}

We first give a recursive construction for the tree amplitudes. Starting from one  ``central'' vertex, we attach four lower point tree amplitudes to it. A sketch of this construction is shown as:
\begin{figure}[H]
\begin{center}
  \includegraphics[width=4cm]{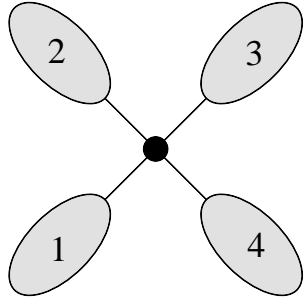}
  \\
  \caption{Constructing a tree amplitude from lower point amplitudes.
  }\label{partition1}
\end{center}
\end{figure}
Let $ P(1..6)$ be the set of all partitions of the set $(1,\ldots,n)$. Then each group of the four tree amplitudes is determined by a certain partition
$\s^{i}=\{\s_1^{i},\s_2^{i},\s_3^{i},\s_4^{i}\} \in P(1..n)$ of the external states $(1,...,n)$ \fn{Note that to avoid over-counting of the partition, we fix the line ``1'' to be in $\s_1^{i}$.}  and contributes
\begin{equation}
  \frac{A_{\s^{i}_{1}}}{S_{\s^{i}_{1}}}\frac{A_{\s^{i}_{2}}}{S_{\s^{i}_{2}}}\frac{A_{\s^{i}_{3}}}{S_{\s^{i}_{3}}}
\frac{A_{\s^{i}_{4}}}{S_{\s^{i}_{4}}}\bigg|_{1 \in \s^{i}_{1}}
\end{equation}
where $A_{\s^{i}_{k}}$ are (physical) $n^{i}_{k}$-pt amplitudes with $\sum_{k=1}^{4}  n^{i}_{k}=n+4$.

Summing over all the partitions gives:
\begin{equation}\label{pretree}
    \sum_{ P(1..6)}
  \frac{A_{\s^{i}_{1}}}{S_{\s^{i}_{1}}}\frac{A_{\s^{i}_{2}}}{S_{\s^{i}_{2}}}\frac{A_{\s^{i}_{3}}}{S_{\s^{i}_{3}}}
\frac{A_{\s^{i}_{4}}}{S_{\s^{i}_{4}}}\bigg|_{1 \in \s^{i}_{1}} \, .
\end{equation}
We also need a symmetry factor due to an over-counting. One can show that each diagram of an $n$-point amplitude in $\phi^4$-theory has $(n-2)/2$ vertices and each vertex could be the central vertex. Hence we see each tree diagram in this construction is counted $V=(n-2)/2$ times. Therefore the tree amplitude is:

\begin{equation}\label{tree1}
   A_{\text{tree}}=\frac{2}{n-2}       \sum_{ P(1..6)}
  \frac{A_{\s^{i}_{1}}}{S_{\s^{i}_{1}}}\frac{A_{\s^{i}_{2}}}{S_{\s^{i}_{2}}}\frac{A_{\s^{i}_{3}}}{S_{\s^{i}_{3}}}
\frac{A_{\s^{i}_{4}}}{S_{\s^{i}_{4}}}\bigg|_{1 \in \s^{i}_{1}} \, .
\end{equation}

Now we proceed to compute the bubble coefficient.  As discussed in section \ref{scalarQFT}, the only non-vanishing contributions after large BCFW shifts come from the following types of cuts:

\begin{figure}[H]
\begin{center}
  \includegraphics[width=5cm]{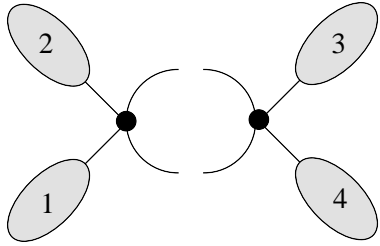}
  \caption{Finite contribution to the bubble coefficient, each of the sub-blob stands for a physical amplitude.}\label{c2structure}
\end{center}
\end{figure}

According to ~\eqref{bbprop}, the bubble coefficient takes the following form\fn{Again we fix the line ``1'' to be in sub-amplitude $\s_{1}^{i}$ to avoid over-counting.}
\begin{equation}\label{scalarc2}
      \cc_{2}=3\a_{4}\times \sum_{ P(1..6)}
  \frac{A_{\s^{i}_{1}}}{S_{\s^{i}_{1}}}\frac{A_{\s^{i}_{2}}}{S_{\s^{i}_{2}}}\frac{A_{\s^{i}_{3}}}{S_{\s^{i}_{3}}}
\frac{A_{\s^{i}_{4}}}{S_{\s^{i}_{4}}}\bigg|_{1 \in \s^{i}_{1}} \, \, \, .
\end{equation}
Crucially, the factor of 3 comes from the fact that there are three cuts, $\s_{1}^{i}\s_{2}^{i}\big|\s_{3}^{i}\s_{4}^{i}$, $\s_{1}^{i}\s_{3}^{i} \big |\s_{2}^{i}\s_{4}^{i}$ and $\s_{1}^{i}\s_{4}^{i} \big |\s_{2}^{i}\s_{3}^{i}$, that have the same partition of the external states and contribute the same to $\cc_{2}$.

Combining \eqref{tree1} and \eqref{scalarc2}, we have
\begin{eqnarray}
\nonumber  \cc_2&=&3\a_{4}\times
\frac{n-2}{2}A_{\text{tree}}=-\b_0 \frac{n-2}{2} \a_{4}A_{\text{tree}} \, ,  \label{finalsum}
\end{eqnarray}
which finally proves eq.~\eqref{treebubb0}, and shows that the bubble coefficient and the tree amplitudes in $\phi^4$ are exactly proportional.  The proportional constant is $3\times \frac{n-2}{2}=-\b_{0} \times V$, which confirms our statement that this coefficient should be related to the UV behavior of the coupling.

\section{Yukawa theory }\label{YukawaQFT}

In this section, we consider Yukawa theory (with complex scalars).
One can again use a Feynman diagram analysis to track the large-$z$ contributions if the diagram has a pure fermion loop. Non-trivial $z$-dependence of external fermion lines becomes cumbersome to handle when both fermions and scalars are present in the loop. For such cases, we use helicity amplitudes in the cut.

\subsection{Yukawa bubbles with Feynman diagrams}

We now consider one-loop scalar amplitudes in Yukawa theory. For scalar loops, the computation is identical with previous section. For fermion loops there are two classes of diagrams, depending on the arrangement of the helicities of the fermions in the loop. According to convention, we treat all momenta as outgoing; helicity is assigned accordingly. In the first case, the fermions on the tree amplitude on one side of the cut, with all momenta outgoing, have the same helicity. In the second case, they have opposite helicity.

\begin{figure}
\begin{center}
\includegraphics[width=4in]{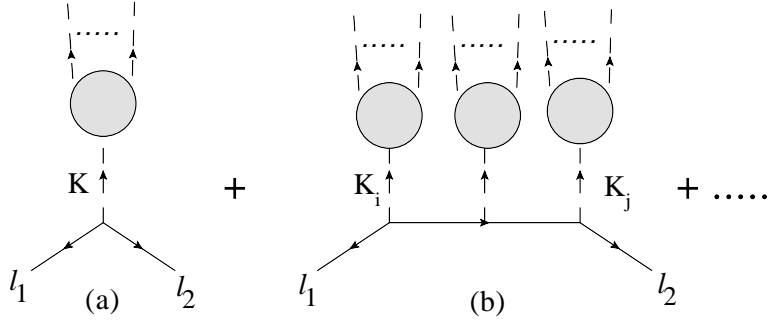}
\caption{Tree diagram with two fermions and all scalars. Here the dashed lines indicate scalars while solid lines are fermions.  Arrows on the solid lines indicate fermion helicity.  If the fermions have the same helicity, with fermion momenta outgoing, the contributing tree diagram must have an even number of fermion propagators. Under the BCFW deformation of $l_1$ and $l_2$, the first diagram is $\mathcal{O}(1)$ as $z\rightarrow\infty$, while the second vanishes as $1/z$.}
\label{2F1S}.
\end{center}
\end{figure}

For the first case, since the fermions on either side of the cut have the same helicity the tree amplitudes on both sides of the cut must have an even number of fermion propagators, as shown in Fig.~\ref{2F1S}. The first diagram is ${\cal O}(z^0)$, since the fermion vertex gives $\langle l_1(z)l_2\rangle=\langle l_1l_2\rangle$,\footnote{The other helicity configuration $[l_1\hat{l}_2]$ would appear on the other side of the cut.} under the shift in eq.~(\ref{bcfwshift}). The second diagram is order ${\cal O}(z^{-1})$, essentially due to the two shifted fermion propagators:
\eqa
{\rm Fig}.\ref{2F1S}(b)&=&\left.\langle \hat{l}_1(z)| \frac{\displaystyle{\not}l_1-\displaystyle{\not}K_i+z\displaystyle{\not}q}{(l_1+K_i - qz)^2}\frac{\displaystyle{\not}l_2+\displaystyle{\not}K_j-\displaystyle{\not}qz}{(l_2+K_j-qz)^2}|l_2\rangle\right|_{z\rightarrow\infty}
\nonumber\\ &=&
\frac{\langle l_2|\displaystyle{\not} K_i\displaystyle{\not} K_j|l_2\rangle}{4z(q\cdot K_i)(q\cdot K_j)}+\mathcal{O}\bigg(\frac{1}{z^2}\bigg) \nonumber
\eqae
where $K_i,K_j$ are the sum of external momenta in each scalar line. Diagrams with more than two propagators which depend on the shifted loop momenta similarly fall off as $1/z$ for large $z$. The same $z$-scaling holds on the opposite side of the cut.  Only the product of the first term in Fig.~\ref{2F1S} on either side of the cut contribute the bubble coefficient, and their phase-space integral is simply:
\eqa
-\frac{1}{2\pi i}\int d{\rm LIPS}\; [l_1l_2]\langle l_1l_2\rangle&=&-\frac{K^2}{2\pi i}\int d{\rm LIPS}*1=K^2\,,
\eqae
where $K$ is the outgoing external momentum. For $K^2 \neq 0$, each scalar line has an associated propagator, $1/K^2$, and the bubble coefficient simply gives $\frac{K^2}{(K^2)^2}=\frac{1}{K^2}$. Such a bubble coefficient then corresponds to a scalar propagator renormalization in the tree diagram.

\begin{figure}
\begin{center}
\includegraphics[width=5.5in]{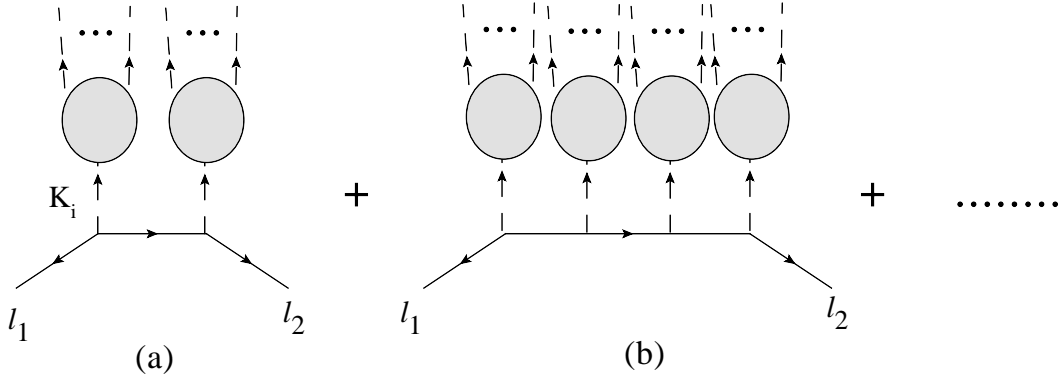}
\caption{If the fermions have opposite helicity, with fermion momenta outgoing, the contributing tree diagrams must have odd number of fermion propagators. Again, arrows on the solid lines indicate fermion helicity.}
\label{2f2s}
\end{center}
\end{figure}
Next we consider fermion loops with opposite helicities on either side of the two-particle cut. The Feynman diagrams appearing on one side of the cut are displayed in Fig.~\ref{2f2s}. The leading large-$z$ behavior is again in the first diagram:
\eq
{\rm Fig}.\ref{2f2s}(a)\{l_1^{-\frac{1}{2}},l_2^{+\frac{1}{2}}\}=\frac{\langle \hat{l}_{1}(z)|K_i| \hat{l}_2(z)]}{(l_1+K_i - qz )^2}\bigg|_{z\rightarrow\infty}=z+ {\cal O}(z^0)\,,
\eqe
where we have indicated the helicities of the loop leg in the curly bracket. The remaining diagrams are sub-leading in $1/z$ and do not contribute. On the other side of the two-particle cut, the same tree diagrams contribute, save with the helicity of $l_1$ and $l_2$ exchanged. The leading large-$z$ behavior again sits in the first diagram. However, this time it scales as $1/z$, since the polarization spinors are not shifted in this helicity configuration. One has for the leading, $1/z$, term
\eq
{\rm Fig}.\ref{2f2s}(a)\{l_1^{+\frac{1}{2}},l_2^{-\frac{1}{2}}\}=\frac{\langle l_{2}|K_i|l_1]}{(l_1+K_i - qz )^2}\bigg|_{z\rightarrow\infty}= \frac{1}{z} + {\cal O}\bigg( \frac{1}{z^2} \bigg) \, . \nonumber
\eqe
The $z^{0}$ piece of the product of the two sides is just $1$. Thus the contribution to the bubble coefficient of this fermion-loop is, again, the tree diagram obtained by replacing the loop by a 4-point contact vertex. Renormalizability then requires the presence of a $|\phi \phi^*|^2$-vertex for the Yukawa theory.

\subsection{Yukawa bubbles from on-shell amplitudes}
For more general processes, it is difficult to isolate the Feynman diagrams that do not vanish at large $z$.
Instead of studying the large $z$ behavior diagram by diagram, here we compute the bubble coefficients for $A_{\<f f \bar{f} \bar{f}\>}^{\rm tree}$ and $A_{\<f, \bar{f}, \f, \f^{*}\>}^{\rm tree}$ using helicity amplitudes. These can be thought of as a sort of warm-up for extracting bubble coefficients in Yang-Mills.

First we list the relevant tree amplitudes of Yukawa theory in Table \ref{Yukawatree}.
\begin{table}[H]
\begin{center}
\begin{tabular}{c|c}
external states   &  amplitudes \\\hline\hline
$\<f(1) f (2)\bar f(3) \bar f(4)\> $ & $g^{2}\frac{\<12\>}{\<34\>}$\\\hline
$\<f(1) \bar f(2) \f(3)\f^{*}(4)\> $ & $g^{2}\frac{\<14\>}{\<24\>}$\\\hline
$\<\f(1) \f(2) \f^{*}(3) \f^{*}(4)\> $ & $\l$\\\hline
\end{tabular}
\end{center}
\caption{On-shell tree amplitudes in Yukawa theory.}
\label{Yukawatree}
\end{table}

\subsubsection{Example 1: 4-fermion bubble coefficient}
To calculate the bubble-coefficient of the 4-fermion loop amplitude, $A_{\<f f \bar{f} \bar{f}\>}^{1-{\rm loop}}$, note that it has three distinct cuts:
\begin{enumerate}

\item The $(f_1 f_2 | \bar{f}_3 \bar{f}_4)$-cut:  here, only the fermion loop contributes and the product of tree amplitudes in the cut are $A^{\rm tree}_{\< f_1f_2\bar{f}_{l_1} \bar{f}_{l_2}\>}A^{\rm tree}_{\<f_{l_2}f_{l_1}\bar{f}_3\bar{f}_4\>}$.  The contribution of this product of tree amplitudes to the overall bubble coefficient is:
\begin{equation}
\frac{1}{(2\pi i)^2}\int d{\rm LIPS} \oint_{\cal C} \frac{dz}{z} \bigg( -\frac{\<12\>}{\<\hat{l}_{1} l_{2}\>}\frac{\<l_{2}\hat{l}_{1}\>}{\<34\>} \bigg) g^{4}= -\frac{1}{2\pi i}\int d{\rm LIPS}\frac{\<12\>}{\<34\>}g^{4} = \frac{\<12\>}{\<34\>} g^4 \ . \nonumber
\end{equation}

\item The $(f_1 \bar{f}_3| f_2 \bar{f}_4)$-cut: here, both the fermion and scalar loops contribute. The contribution from the fermion loop reads,
\begin{equation}
\nonumber\frac{1}{(2\pi i)^2}\int d{\rm LIPS} \oint_{\mathcal{C}} \frac{dz}{z} \bigg(-\frac{\<1\hat{l}_{1}\>}{\<3 l_{2}\>}\frac{\<2 l_{2}\>}{\<4 \hat{l}_{1}\>} \bigg) g^{4} = \frac{1}{2\pi i} \int d{\rm LIPS} \frac{\< 1 l_2 \> \< 2 l_2 \>}{\< 3 l_2 \> \< 4 l_2 \>} g^4\, , \nonumber
\end{equation}
where we have expanded
\eq
\nonumber\frac{\<1\hat{l}_{1}\>}{\<4 \hat{l}_{1}\>}\bigg|_{z\rightarrow\infty}=\frac{\<1l_2\>}{\<4 l_2\>}+\mathcal{O}\bigg(\frac{1}{z}\bigg)\,.
\eqe
To evaluate the $d{\rm LIPS}$ integral, we rewrite it as a total derivative in the anti-holomorphic spinor $\tilde{\lambda}$. Being a total derivative, the integration is then localized by the holomorphic anomalies generated by $\<\lambda, i\>$ in the denominator. We refer to appendix \ref{Holomorphic} for detailed discussion. Using eq.~(\ref{derivative}), we can write the $d{\rm LIPS}$ integral as
\eqa
\nonumber &&\frac{1}{(2\pi i)^2}\int d{\rm LIPS} \frac{\< 1 l_2 \> \< 2 l_2 \>}{\< 3 l_2 \> \< 4 l_2 \>} g^4\\
\nonumber&=&-g^4\frac{1}{2\pi i}\int \<\l d\l\>d\tilde{\l}^{\dot\alpha}\frac{\partial}{\partial\tilde{\l}^{\dot\a}}\bigg[\frac{\<\a|P_{1,3}|\tilde\l]}{\<\l|P_{1,3}|\tilde\l]\<\l\a\>}\frac{\< 1 \l \> \< 2 \l \>}{\< 3 \l \> \< 4 \l \>}\bigg]=g^4\frac{\<12\>}{\<34\>}\,.
\eqae
To arrive at the final result, one can conveniently choose $\alpha=1$, and since the residue for the pole $\< 3 \l \>$ is zero with $\alpha=1$, the only pole in the denominator that localizes the integral is $\< 4 \l \>$. The contribution from the scalar loop can be computed in a similar fashion and gives $-2g^4\frac{\<12\>}{\<34\>}$, where the $2$ comes from the intermediate state sum. Thus the bubble coefficient for this cut is $-g^4\frac{\<12\>}{\<34\>}$.

\item The $(f_1 \bar{f}_4| f_2 \bar{f}_3)$-cut.  This is a relabeling of the $(f_1 \bar{f}_3| f_2 \bar{f}_4)$-cut, and we get $-\frac{\<12\>}{\<34\>}g^{4}$.

\end{enumerate}
Summing over all channels gives the sum of bubble coefficients for $1$-loop $\<f f\bar f\bar f\> $ amplitude:
\begin{equation}
\cc_{2}^{\<f f\bar f\bar f\> }=(1-1-1)\frac{\<12\>}{\<34\>}g^{4}=-g^{2}A^{\text{tree}}_{\<f f\bar f\bar f\> } \ .
\end{equation}
Thus we see that the proportionality between the sum of bubble coefficients and the tree amplitude is achieved without the need to introduce a new 4-fermion vertex.

\subsubsection{Example 2: two-fermion two-scalar bubble coefficient}
The second example is $A_{\<f \bar f \f \f^{*}\>}^{1-{\rm loop}}$ which also has three distinct cuts:
\begin{enumerate}

\item The $(f_1 \bar{f}_2| \f_3 \f^*_4)$-cut:  both scalars and fermions run in the loop.  Large-$z$ poles for either contribution have identical $d{\rm LIPS}$ integrands.  Direct computation shows they vanish.

\item The $(f_1 \f_3| \bar{f}_2 \f^*_4)$-cut: in this case, the contribution comes from a fermion in one and a scalar in the other cut loop-line. The product of the two tree amplitudes are:
$$\bigg(\frac{\<1 l_{1}\>\<l_{2}4\>}{\<l_{2} l_{1}\>\<2 4 \>}+\frac{\<1 l_{2}\>\<l_{1}4\>}{\<l_{1} l_{2}\>\<2 4 \>}\bigg)g^{4}=-\frac{\<1 4 \>}{\<2 4 \>}g^{4}\,,$$
where we have used the Schouten identity. The $d{\rm LIPS}$ integrand is independent of the loop momenta, hence the bubble coefficient is simply $-\frac{\<1 4 \>}{\<2 4 \>}g^{4}$.

\item The $(f_1 \f^*_4| \bar{f}_2 \f_3)$-cut.  The double-cut has a trivial residue for large $z$, and thus does not contribute to the bubble coefficient.
\end{enumerate}
Thus the bubble coefficient of  $\<f \bar f \f \f^{*}\> $ is:
\begin{equation}\label{2fermions2scalar}
\cc_{2}^{\<f \bar f \f \f^{*}\> }=-\frac{\<14\>}{\<24\>}g^{4}=-g^{2}A^{\text{tree}}_{\<f \bar f \f \f^{*}\> } \ .
\end{equation}

In the above two examples, the sum of bubble coefficients is proportional to the tree amplitude, so one-loop renormalizability does not require us to introduce interaction terms $\psi^4$ or $\psi^2\phi^2$ in the effective action.
This is, of course, what we expected as these are higher-dimensional non-renormalizable interactions.

\section{Sum of MHV bubble coefficients for pure Yang-Mills\label{NAMHV}}
As mentioned in subsection~\ref{pYM}, the observed structure of cancellations for $\mathcal{N}=1,2$ super Yang-Mills theory is present in pure Yang-Mills as well. However, it is more involved to derive this since the $\mathcal{O}(z^0)$ part of the BCFW-shifted two-particle cut contains higher-order collinear poles. Nevertheless, adjacent channels again share these higher-order CCP, and their contribution to the sum of bubble coefficient also cancels. The cancellation of CCP renders the summation down to the terminal poles, which evaluate to $11/6A_n^{\rm tree}$.  Here, we explicitly deal with the details of this computation.

\begin{figure}
\begin{center}
\includegraphics[width=2in]{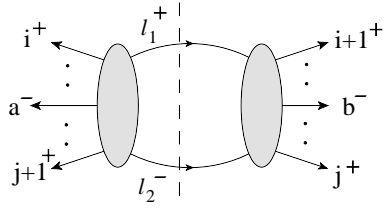}
\caption{The $(l_1^{+},l_2^{-})$ helicity configuration for the two-particle cut $(j+1,.., a,..,i|i+1,..,b,..,j)$ of $A_n^{\rm MHV}(a^{-},b^{-})$.   }
\label{GeneralCut2}
\end{center}
\end{figure}

We begin with a generic BCFW-shifted two-particle cut of $(j+1,.., a,..,i|i+1,..,b,..,j)$ for non-adjacent MHV amplitude $A_n^{\rm MHV}(a^{-},b^{-})$. Choosing the helicity configuration for the internal lines to be $(l_1^{+},l_2^{-})$ on the LHS of the cut, as shown in Fig.~\ref{GeneralCut2}, one has:
\begin{eqnarray}
S^{(i,j)}_{a,b} &=&  A_n^{\rm tree} \bigg\{ \frac{\< i, i+1 \>\< b, \hat{l}_1 \>}{\< i, \hat{l}_1\> \< \hat{l}_1, i+1\>} \bigg\}\bigg\{ \frac{\< j, j+1 \>\< a, l_2 \>}{\< j,l_2 \>\< l_2, j+1 \>}\bigg\}\frac{ \<l_2,l_1\>}{\< a, b \>} \bigg( \frac{\< a, l_2 \> \< b, \hat{l}_1 \>}{\< a, b \>\<l_1,l_2\>} \bigg)^3 \, \, \, . \nonumber\\
\label{PoleShow1}
\end{eqnarray}
Let us extract the bubble coefficient by shifting the loop legs as $| \hat{l}_1\>\rightarrow |l_1\>+ z|l_2\>$. Note that for our choice of shift, it will be convenient to take $|l_2\>=|\lambda\>$ as the $d{\rm LIPS}$ integration spinor. Under the $d{\rm LIPS}$ integration, there are three kinds of poles that would contribute to the holomorphic anomaly: (1) the $1/\<\lambda\alpha\>$ poles that arise from writing the $d{\rm LIPS}$ integral as a total derivative, (2) the collinear poles of the form $1/\<l_2i\>$ and (3) the poles that come from expanding $1/\<\hat{l}_1i\>$ in $1/z$ to obtain the $\mathcal{O}(z^0)$ piece at $z\rightarrow\infty$.

We can remove the poles of type (1) by choosing $|\alpha\>=|a\>$, since the factor of  $\< l_2, a\>$ in the numerator of eq.~(\ref{PoleShow1}) will cancel this pole. Thus the only contributions remaining are of type (2) and (3). We rewrite eq.~(\ref{PoleShow1}) such that each type of pole is separated:
\begin{eqnarray}
\nonumber S^{(i,j)}_{a,b}&=&  A_n^{\rm tree}\frac{\< b, \hat{l}_1 \>^{3}}{\<l_1,l_2\>^{2}}\bigg\{ \frac{\<i,b\>}{\<\hat{l}_1,i\> }-\frac{\<i+1,b\>}{\<\hat{l}_1,i+1\>}\bigg\}\bigg\{ \frac{\<j, a\>}{\<l_2, j\>}-\frac{\<j+1, a\>}{\<l_2, j+1\>}\bigg\} \bigg(\frac{-\< a, l_2 \>^{3} }{\< a, b \>^{4}} \bigg)\, .
\end{eqnarray}
Next, we expand around $z\rightarrow\infty$ obtaining,
\begin{eqnarray}
\nonumber \cs^{(i,j)}_{a,b}(\l)\equiv S^{(i,j)}_{a,b}\big|_{\co(z^{0})}\!\!\!\!&=&\!\!\!A_n^{\rm tree}\bigg\{ \cg^{(i,j)}_{a,b,i}(\l)-\cg^{(i,j)}_{a,b,i+1}(\l)\bigg\}\bigg\{ \frac{\<j, a\>}{\<\l, j\>}-\frac{\<j+1, a\>}{\<\l, j+1\>}\bigg\} \bigg(\frac{-\< a, \l \>^{3} }{\< a, b \>^{4}} \bigg)\,, \\
&=&A_n^{\rm tree}\big\{ I\big\}\big\{ J \big\} \bigg(\frac{-\< a, \l \>^{3} }{\< a, b \>^{4}} \bigg)\label{PoleShow4}
\end{eqnarray}
where we've used $\big\{ I\big\}$ and $\big\{ J\big\}$ as a short hand notation for the terms in the curly bracket. Later we will see manifest cancellation of CCP for the terms in $\big\{ I\big\}$ under the summation over the $i$-indices, and similarly for $\big\{ J\big\}$ under the summation over the $j$-indices. The new functional ${\mathcal G}^{(i,j)}_{a,b,i}$ is defined as:

\eqa
\nonumber {\mathcal G}^{(i,j)}_{a,b,i}\!\!&\equiv&\!\!\!\left.\frac{\<i,b\>}{\<l_1,l_2\>^{2}}\frac{\<b,\hat{l}_1\>^3}{\<\hat{l}_1,i\> }\right|_{\mathcal{O}(1)}\!\!\!\!=\frac{\<i,b\>}{\<l_1,l_2\>^{2}}\left(3\frac{\<b,l_1\>^2\<b,l_2\>}{\<l_2,i\>}-3\frac{\<b,l_1\>\<b,l_2\>^2\<l_1,i\>}{\<l_2,i\>^2}+\frac{\<b,l_2\>^3\<l_1,i\>^2}{\<l_2,i\>^3}\right)\\
&=&\!\!\!\frac{\<i,b\>}{\<\l|P_{i,j}|\l]^{2}}\left(3\frac{\<b|P_{i,j}|\l]^2\<b,\l\>}{\<\l,i\>}+3\frac{\<b|P_{i,j}|\l]\<b,\l\>^2\<i|P_{i,j}|\l]}{\<\l,i\>^2}+\frac{\<b,\l\>^3\<i|P_{i,j}|\l]^2}{\<\l,i\>^3}\right)
\label{gabi}
\eqae
and similarly,
\eqa
\nonumber {\mathcal G}^{(i,j)}_{a,b,i+1}\!\!&\equiv&\!\!\!\frac{\<i+1,b\>}{\<\l|P_{i,j}|\l]^{2}}\left(3\frac{\<b|P_{i,j}|\l]^2\<b,\l\>}{\<\l,i+1\>}+3\frac{\<b|P_{i,j}|\l]\<b,\l\>^2\<i+1|P_{i,j}|\l]}{\<\l,i+1\>^2}+\frac{\<b,\l\>^3\<i+!|P_{i,j}|\l]^2}{\<\l,i+1\>^3}\right)\,.\\
\label{gabi1}
\eqae
Note the function $ \cg^{i,j}_{a,b,i}(\l)$ has higher-order (aka not simple) collinear poles in $\<\l,i\>$, which will require extra care in using the holomorphic anomaly as we later discuss.

The $d{\rm LIPS}$ integral of \eqref{PoleShow4} is localized by the four poles appearing in the curly brackets, and it will be convenient to separate the contributions from the first and second curly brackets. Writing,
\begin{equation}
\sum\limits_{i,j} \frac{-1}{2\pi i}\int d{\rm LIPS}\;\; \cs^{(i,j)}_{a,b}(\l)=\sum\limits_{i,j} \frac{-1}{2\pi i}\int d{\rm LIPS}\;\; \left[\cs^{(i,j)}_{a,b}(\l)\bigg|_{\{J\}}+\cs^{(i,j)}_{a,b}(\l)\bigg|_{\{I\}}\right]\,,
 \end{equation}
where $|_{\{I\}}$ indicates the contributions that arises from the presence of poles in $\{I\}$. We first consider the sum of residues of the simple poles $1/\<\l,j\>$ and $1/\<\l,j+1\>$ in the second curly bracket:
\begin{eqnarray}
\nonumber \!\!\!\!\!\sum\limits_{i,j} \cs^{(i,j)}_{a,b}(\l)\bigg|_{\{J\}} \!\!\!\!&=&\sum\limits_{i,j} A_n^{\rm tree}\bigg\{ \cg^{(i,j)}_{a,b,i}(\l)-\cg^{(i,j)}_{a,b,i+1}(\l)\bigg\}\bigg\{ \frac{\<j, a\>}{\<\l, j\>}\bigg\} \bigg(\frac{-\< a, \l \>^{3} }{\< a, b \>^{4}} \bigg)\bigg|_{\<l, j\>}\\
\nonumber &&-\sum\limits_{i,j} A_n^{\rm tree}\bigg\{ \cg^{(i,j)}_{a,b,i}(\l)-\cg^{(i,j)}_{a,b,i+1}(\l)\bigg\}\bigg\{ \frac{\<j+1, a\>}{\<\l, j+1\>}\bigg\} \bigg(\frac{-\< a, \l \>^{3} }{\< a, b \>^{4}} \bigg)\bigg|_{\<\l, j+1\>}\\
\nonumber &=&\sum\limits_{i,j} A_n^{\rm tree}\bigg\{ \cg^{(i,j)}_{a,b,i}(\l)-\cg^{(i,j)}_{a,b,i+1}(\l)\bigg\}\bigg\{ \frac{\<j, a\>}{\<\l, j\>}\bigg\} \bigg(\frac{-\< a, \l \>^{3} }{\< a, b \>^{4}} \bigg)\bigg|_{\<\l, j\>}\\
\nonumber  &&-\sum\limits_{i,j'} A_n^{\rm tree}\bigg\{ \cg^{(i,j'-1)}_{a,b,i}(\l)-\cg^{(i,j'-1)}_{a,b,i+1}(\l)\bigg\}\bigg\{ \frac{\<j', a\>}{\<\l, j'\>}\bigg\} \bigg(\frac{-\< a, \l \>^{3} }{\< a, b \>^{4}} \bigg)\bigg|_{\<\l, j'\>}\,,\\
 \label{sumj}
\end{eqnarray}
where we've used $|_{\<\l, j\>}$ to indicate the collinear pole on which the integrand will be localized. From \eqref{gabi}, we see that $\cg^{(i,j)}_{a,b,i}(\l)=\cg^{(i,j-1)}_{a,b,i}(\l)$ when localized at $\l\to j$.\fn{This again can be seen from the fact that on the pole, $P_{i,j}|j]=P_{i,j-1}|j]$.} Treating $j$ as and $j'$ as dummy variables, the two summation simply cancels with each other and one is left with zero! Of course this is the wrong result and the subtlety lies in the summation limits. We will discuss the limits in detail in the next subsection. For now, we will show the same cancellation occurs for the higher-order poles in the first curly bracket.

The contributions from the poles in the first curly bracket in \eqref{PoleShow4} can be written as:

\begin{eqnarray}
\nonumber \!\!\!\!\!\sum\limits_{i,j} \cs^{(i,j)}_{a,b}(\l)\bigg|_{\{I\}} \!\!\!\!&=&\sum\limits_{j,i} A_n^{\rm tree}\bigg\{ \cg^{(i,j)}_{a,b,i}(\l)\bigg\}\bigg\{ \frac{\<j, a\>}{\<\l, j\>}-\frac{\<j+1, a\>}{\<\l, j+1\>}\bigg\} \bigg(\frac{-\< a, \l \>^{3} }{\< a, b \>^{4}} \bigg)\bigg|_{\<\l, i\>}\\
\nonumber &&-\sum\limits_{j,i} A_n^{\rm tree}\bigg\{ \cg^{(i,j)}_{a,b,i+1}(\l)\bigg\}\bigg\{ \frac{\<j, a\>}{\<\l, j\>}-\frac{\<j+1, a\>}{\<\l, j+1\>}\bigg\} \bigg(\frac{-\< a, \l \>^{3} }{\< a, b \>^{4}} \bigg)\bigg|_{\<\l, i+1\>}\\
\nonumber &=&\sum\limits_{j,i} A_n^{\rm tree}\bigg\{ \cg^{(i,j)}_{a,b,i}(\l)\bigg\}\bigg\{ \frac{\<j, a\>}{\<\l, j\>}-\frac{\<j+1, a\>}{\<\l, j+1\>}\bigg\} \bigg(\frac{-\< a, \l \>^{3} }{\< a, b \>^{4}} \bigg)\bigg|_{\<\l, i\>}\\
&&-\sum\limits_{j,i'} A_n^{\rm tree}\bigg\{ \cg^{(i'-1,j)}_{a,b,i'}(\l)\bigg\}\bigg\{ \frac{\<j, a\>}{\<\l, j\>}-\frac{\<j+1, a\>}{\<\l, j+1\>}\bigg\} \bigg(\frac{-\< a, \l \>^{3} }{\< a, b \>^{4}} \bigg)\bigg|_{\<\l, i'\>} .
\label{Useful}
\end{eqnarray}
Here, the integral will be localized by the poles that are present in $\cg^{(i,j)}_{a,b,i}(\l)$, which besides simple poles, has higher-order poles at the same kinematic point. Repeated use of the Schouten identity allows one to extra the contribution of the higher-order poles to the simple pole~ \cite{HigherPoles}, which we review in appendix~\ref{Holomorphic}. Denoting the resulting expression as $\ch^{(i,j)}_{a,b,i}(\l)$, we have
\begin{eqnarray}
\nonumber\!\!\!\!\!\sum\limits_{i,j} \cs^{(i,j)}_{a,b}(\l)\bigg|_{\{I\}} \!\!\!\!&=&\sum\limits_{j,i}\frac{ A_n^{\rm tree}}{\< a, b \>^{4}}\bigg\{ \ch^{(i,j)}_{a,b,i}(\l)-\ch^{(i,j+1)}_{a,b,i}(\l)-\ch^{(i-1,j)}_{a,b,i}(\l)+\ch^{(i-1,j+1)}_{a,b,i}(\l)\bigg\}\bigg|_{\<\l, i\>}\,.\\
\label{sumi}
\end{eqnarray}
The explicit form of $\ch^{(i,j)}_{a,b,i}(\l)$ is given in eq.~(\ref{Comeon}). The key fact of $\ch^{(i,j)}_{a,b,i}(\l)$ and $\ch^{(i-1,j)}_{a,b,i}(\l)$ is that they become identical when the integrand is evaluated on the pole $1/\<\lambda ,i\>$ and integrated on the real contour $\tilde\lambda=\bar\lambda$. Therefore eq.~(\ref{sumi}) again gives zero!

While the limits of the summation requires careful treatment, our analysis shows that indeed for non-adjacent MHV, the cancellation of CCP again reduces the sum of bubble coefficients to a few terminal terms which we will now identify.

\subsection{$d{\rm LIPS}$ integrals of higher-order poles}

As we have seen, generic pole terms in non-adjacent MHV bubble coefficients' $g(\l)$s generically have higher-order poles.  We evaluate the integrands in a manner following that in section 2.3 of \cite{HigherPoles}.  Specifically, the $d{\rm LIPS}$ integrands are rational functions of $\l$ of degree $-2$.  We can recursively reduce the degree of $\l$ in the numerator and denominator by one unit each, through repeated application of the following Schouten identity \cite{HigherPoles}:
\eq\label{speschouten}
\frac{\< a, \l \>}{\< \b, \l \> \< \g, \l \>} = \frac{\<a, \b \>}{\< \g, \b \>} \frac{1}{\< \b, \l \>} + \frac{\< a, \g \>}{\< \b, \g \>} \frac{1}{\< \g, \l \>} \ .
\eqe
Repeated application reduces integrands with higher-order poles to sums of integrands with either simple poles, or to multiple poles, such as $1/\<a, \l \>^{2}, \ \<x, \l \>/ \<a, \l\>^{3}$ etc., with trivial residues as $| \l \> \rightarrow | a \>$.  The generic form for the residues at second- and third-order poles are:
\eqa
\frac{1}{\< \b, \l \>^2} \prod_{i = 1}^{n} \frac{\< a_i, \l \>}{\< b_i, \l \>} \bigg|_{\<\l, \b\>} &=&  \prod_{i = 1}^{n} \frac{\< a_i, \b \>}{\< b_i, \b \>}  \sum_{1 \leq i \leq n} \frac{\< a_i, b_i \>}{\<a_i, \b \> \< b_i, \b \>} \ ,
\label{reshp1}\\
\frac{\< \xi, \l \>}{\< \b, \l \>^3} \prod_{i = 1}^{n} \frac{\< a_i, \l \>}{\< b_i, \l \>}\bigg|_{\<\l, \b\>} &= &\< \xi, \b \>  \prod_{i = 1}^{n} \frac{\< a_i, \b \>}{\< b_i, \b \>}  \bigg\{ \sum_{1 \leq i \leq j \leq n  } \frac{\< a_i, b_i \>}{\<a_i, \b \> \< b_i, \b \>} \frac{\< a_j, b_j \>}{\<a_j, \b \> \< b_j, \b \>} \label{reshp2}\\
& & \, \, \, \, \, \, \, \, \, \, \, \, \, \, \, \, \, \, \, \, \, \, \, \, \, \,  \, \, \, \, \, \, \, \, \, \, \, \, \, \, \, \, \,   + \sum_{1\leq k \leq n} \frac{\< a_k, b_k \>}{\< a_k, \b \> \< b_k, \b \>} \frac{\< a_k, \xi \>}{\< \xi, \b \> \< a_k, \b \>} \bigg\} \, \, .  \nonumber
\eqae
With this, we can factor out the irrelevant multiple poles in any expression, for example we have the following rewriting:
\begin{eqnarray}
\frac{\<b,\l\>^2\< \l, a \>^{3} }{\<\l,i\>^2\<\l, j\>\<\l|P_{i,j}|\l]^{4}}\bigg|_{\<\l,i\>}&=&-\frac{\langle i,a\rangle ^3 \langle i,b\rangle ^2 \left(\frac{2 \langle a|P_{i,j}|i] }{\langle i|P_{i,j}|i]  \langle i,a\rangle }+\frac{2 \langle b|P_{i,j}|i] }{\langle i|P_{i,j}|i]  \langle i,b\rangle }+\frac{\langle a,j\rangle }{\langle i,a\rangle  \langle i,j\rangle }\right)}{\langle i|P_{i,j}|i] ^4 \langle i,j\rangle }\,,\label{doublepole}\\
[3mm]
\nonumber \frac{\< \l, a \>^{3}\<b,\l\>^3 }{\<\l, j\>\<\l,i\>^3\<\l|P_{i,j}|\l]^{4}}\bigg|_{\<\l, i\>}&=& \frac{\langle a,i\rangle ^3 \langle b,i\rangle ^3}{\langle i|P_{i,j}|i] ^4 \langle i,j\rangle } \bigg(\frac{3 \langle a|P_{i,j}|i] ^2}{\langle i|P_{i,j}|i] ^2 \langle a,i\rangle ^2}+\frac{3 \langle b|P_{i,j}|i] ^2}{\langle i|P_{i,j}|i] ^2 \langle b,i\rangle ^2}+\frac{2 \langle a|P_{i,j}|i]  \langle a,b\rangle }{\langle i|P_{i,j}|i]  \langle a,i\rangle ^2 \langle b,i\rangle }\\
\nonumber &&+\frac{4 \langle a|P_{i,j}|i\rangle  \langle b|P_{i,j}|i] }{\langle i|P_{i,j}|i] ^2 \langle a,i\rangle  \langle b,i\rangle }+\frac{\langle a,j\rangle ^2}{\langle a,i\rangle ^2 \langle i,j\rangle ^2}+\frac{2 \langle a|P_{i,j}|i]  \langle a,j\rangle }{\langle i|P_{i,j}|i]  \langle a,i\rangle ^2 \langle i,j\rangle }\\
&&+\frac{\langle a,b\rangle  \langle a,j\rangle }{\langle a,i\rangle ^2 \langle b,i\rangle  \langle i,j\rangle }+\frac{2 \langle b|P_{i,j}|i]  \langle a,j\rangle }{\langle i|P_{i,j}|i]  \langle a,i\rangle  \langle b,i\rangle  \langle i,j\rangle }\bigg) \, \, . \label{triplepole}
\end{eqnarray}
Combining these results, we can rewrite $\cg^{(i,j)}_{a,b,i}(\l)\frac{\< \l, a \>^{3} \<j,a\>}{\<\l|P_{i,j}|\l]^2\<\l, j\>}$ to $\ch^{(i,j)}_{a,b,i}$ as in \eqref{sumi}:
\begin{eqnarray}
\nonumber\cg^{(i,j)}_{a,b,i}(\l)\frac{\< \l, a \>^{3} \<j,a\>}{\<\l|P_{i,j}|\l]^2\<\l, j\>}\bigg|_{\<\l, i\>}&=&\frac{\< i, a \>^{3} \<j,a\>}{\<i,j\>}\<i,b\>\bigg(3\frac{\<b|P_{i,j}|i]^2\<b,i\>}{\<\l|P_{i,j}|\l]^2}\\
\nonumber&&+3\<j,a\>\<b|P_{i,j}|i] \<i|P_{i,j}|i]\times \eqref{doublepole}\\
&&+\<j,a\>\<i|P_{i,j}|i]^2\times\eqref{triplepole}\bigg) \, \, .
\label{Comeon}
\end{eqnarray}
The upshot is that \eqref{Comeon} has vanishing residue at the poles $\l=a$ and $\l=b$ . Note that if one considers $\ch^{(i-1,j)}_{a,b,i}$, the only difference is substituting $P_{i,j}$ in $\ch^{(i,j)}_{a,b,i}$ with $P_{i-1,j}$. It can be easily seen that on the pole $1/\<\l,i\>$, the two are equivalent.

\subsection{Terminal poles, terminal cuts and their evaluation}
In determining the limits of the summation, one has to avoid configurations where there is a three point amplitude one side of the cut, as these produce massless bubbles that are set to zero in dimensional regularization. This implies that in the summation of $i,j$ in eq.~(\ref{sumj}) and eq.~(\ref{sumi}), the summation limit of one index will depend on the value of the other.

For eq.~(\ref{sumj}) one sums over the index $j$ first, and the limit is given as:
\begin{equation}\label{junshiftedsum}
\sum\limits_{i,j}=\sum\limits_{i=a+1}^{b-2}\sum\limits_{j=b}^{a-1}+\sum\limits_{j=b+1}^{a-1}\bigg|_{\substack{i=b-1}}+\sum\limits_{j=b}^{a-2}\bigg|_{i=a}\,,
\end{equation}
where $|_{\substack{i=b-1}}$ indicates the index $i$ is held fixed to be $b-1$. Using $j'=j+1$, the summation limit for $j'$ is given as:
\begin{equation}\label{jshiftedsum}
\sum\limits_{i,j'=j+1}=\sum\limits_{i=a+1}^{b-2}\sum\limits_{j'=b+1}^{a}+\sum\limits_{j'=b+2}^{a}\bigg|_{i=b-1}+\sum\limits_{j'=b+1}^{a-1}\bigg|_{i=a}\,.
\end{equation}
Looking back at eq.~\eqref{sumj} we see that there are mismatches in the limits between to two sums, and hence the cancellation is not complete, leaving behind:
\begin{equation}
\sum\limits_{i=a+1}^{b-2}X^{(i,j)}_{a,b}\bigg|_{j=b}-\sum\limits_{i=a+1}^{b-2}X^{(i,j')}_{a,b}\bigg|_{j'=a}+X^{(i,j)}_{a,b}\bigg|_{\substack{i=b-1\\j=b+1}}-X^{(i,j')}_{a,b}\bigg|_{\substack{i=b-1\\j'=a}}+X^{(i,j)}_{a,b}\bigg|_{\substack{i=a\\j=b}}-X^{(i,j')}_{a,b}\bigg|_{\substack{i=a\\j'=a-1}}\,
\label{ResLim}
\end{equation}
where $X^{(i,j)}_{a,b}=-A_n^{\rm tree}\{ \cg^{(i,j)}_{a,b,i}(\l)-\cg^{(i,j)}_{a,b,i+1}(\l)\} \frac{\<j, a\>}{\<\l, j\>}\frac{\< a, \l \>^{3} }{\< a, b \>^{4}}|_{\<\l, j\>}$. The first two terms in eq.~(\ref{ResLim}) evaluates to zero. To see this note that these two sums are evaluated on the pole $1/\<\l, b\>$ and $1/\<\l, a\>$ respectively. Looking at the summand in eq.~(\ref{sumj}) there is a factor $\<\l,a\>$ in the numerator while $\cg^{(i,j)}_{a,b,i}$ has at least one $\<\l,b\>$ in the numerator, as can be seen from eq.~(\ref{gabi}). For the same reason, the fourth and fifth term vanishes as well. The remaining terms are given by:
\begin{eqnarray}
\nonumber\sum\limits_{i,j} \cs^{(i,j)}_{a,b}(\l)\bigg|_{\{J\}} &=&A_n^{\rm tree}\bigg\{ \cg^{(b-1,b+1)}_{a,b,b-1}(\l)-\cg^{(b-1,b+1)}_{a,b,b}(\l)\bigg\}\bigg\{ \frac{\<b+1, a\>}{\<\l, b+1\>}\bigg\} \bigg(\frac{-\< a, \l \>^{3} }{\< a, b \>^{4}} \bigg)\bigg|_{\substack{\<\l, b+1\>}}\\
\nonumber&&-A_n^{\rm tree}\bigg\{ \cg^{(a,a-2)}_{a,b,a}(\l)-\cg^{(a,a-2)}_{a,b,a+1}(\l)\bigg\}\bigg\{ \frac{\<a-1, a\>}{\<\l, a-1\>}\bigg\} \bigg(\frac{-\< a, \l \>^{3} }{\< a, b \>^{4}} \bigg)\bigg|_{\substack{\<\l, a-1\>}}\\
\label{Jfn}
\end{eqnarray}
Therefore, we see the complicated summation \eqref{sumj} reduces to only two terms: the residue at the pole $\<\l,a-1\>=0$ in channel $(i=a,j=a-2)$, and the residue of the pole $\<\l,b+1\>=0$ in channel $(i=b-1,j=b+1)$.

We now look at eq.~(\ref{sumi}), where the index $i$ was summed first. The summation limit is given by:
\begin{equation}\label{iunshiftedsum}
\sum\limits_{j,i}=\sum\limits_{j=b+1}^{a-2}\sum\limits_{i=a}^{b-1}+\sum\limits_{i=a+1}^{b-1}\bigg|_{j=a-1}+\sum\limits_{i=a}^{b-2}\bigg|_{j=b} \, .
\end{equation}
Recalling that $i'=i+1$, the summation limit for $i'$ is given by:
\begin{equation}\label{ishiftedsum}
\sum\limits_{j,i'}=\sum\limits_{j=b+1}^{a-2}\sum\limits_{i'=a+1}^{b}+\sum\limits_{i'=a+2}^{b}\bigg|_{j=a-1}+\sum\limits_{i'=a+1}^{b-1}\bigg|_{j=b}\,.
\end{equation}
Again the mismatch of the summation limits for $i$ and $i'$ leads to uncancelled terms in eq.~\eqref{sumi}, given by:
\begin{equation}
\sum\limits_{j=b+1}^{a-2}Y^{(i,j)}_{a,b}\bigg|_{i=a}-\sum\limits_{j=b+1}^{a-2}Y^{(i',j)}_{a,b}\bigg|_{i'=b}+Y^{(i,j)}_{a,b}\bigg|_{\substack{i=a+1\\j=a-1}}-Y^{(i',j)}_{a,b}\bigg|_{\substack{i'=b\\j=a-1}}+Y^{(i,j)}_{a,b}\bigg|_{\substack{i=a\\j=b}}-Y^{(i',j)}_{a,b}\bigg|_{\substack{i'=b-1\\j=b}}
\label{ResLim2}
\end{equation}
where $Y^{(i,j)}_{a,b}=A_n^{\rm tree}\{ \ch^{(i,j)}_{a,b,i}(\l)-\ch^{(i,j+1)}_{a,b,i}(\l)-\ch^{(i-1,j)}_{a,b,i}(\l)+\ch^{(i-1,j+1)}_{a,b,i}(\l)\}/\< a, b \>^{4}|_{\<\l, i\>}$. The second and fourth term in eq.~(\ref{ResLim2}) evaluates to zero since it has vanishing residue on the pole $1/\<\lambda,b\>$, as can be seen from the presence of $\<\lambda,b\>$ in the numerator of eq.~(\ref{gabi}) and eq.~(\ref{gabi1}). The first and fifth term also vanishes due to the $\<\l,a\>^3$ in the numerator of eq.~(\ref{Useful}). As a result, the sum in eq.~(\ref{sumi}) reduces to
\begin{eqnarray}\label{Ifn}
\nonumber\sum\limits_{i,j} \cs^{(i,j)}_{a,b}(\l)\bigg|_{\{I\}}=\frac{ A_n^{\rm tree}}{\< a, b \>^{4}}\bigg\{\bigg(\ch^{( a+1,a-1)}_{a,b, a+1}(\l)-\ch^{( a+1,a)}_{a,b, a+1}(\l)\bigg)\bigg|_{\substack{\<\l, a+1\>}}\\
-\bigg(\ch^{(b-2,b)}_{a,b,b-1}(\l)-\ch^{(b-2,b+1)}_{a,b,b-1}(\l)\bigg)\bigg|_{\substack{\<\l, b-1\>}}\bigg\}
\end{eqnarray}
Thus the sum localizes to the pole $\<\l,a+1\>=0$ in channel $(i=a+1,j=a-1)$ and $\<\l,b-1\>=0$ in channel $(i=i'-1=b-2,j=b)$.

Collecting all the pieces we now have:
\begin{eqnarray}
\nonumber \cc_{2}(l_1^+,l_2^-)&=&\frac{-1}{2\pi i}\int_{\bar\lambda=\tilde\lambda}d{\rm LIPS}\;\; \frac{A_n^{\rm tree}}{\<a,b\>^4}\bigg\{ -\bigg( \cg^{(b-1,b+1)}_{a,b,b-1}(\l)-\cg^{(b-1,b+1)}_{a,b,b}(\l)\bigg) \frac{\<b+1, a\>}{\<\l, b+1\>}\< a, \l \>^{3}\bigg|_{\<\l, b+1\>}\\
\nonumber&~&+\bigg( \cg^{(a,a-1)}_{a,b,a}(\l)-\cg^{(a,a-1)}_{a,b,a+1}(\l)\bigg)\frac{\<a-1, a\>}{\<\l, a-1\>}\< a, \l \>^{3}\bigg|_{\<\l, a-1\>}\\
\nonumber&~&\bigg(\ch^{( a+1,a-1)}_{a,b, a+1}(\l)-\ch^{( a+1,a)}_{a,b, a+1}(\l)\bigg)\bigg|_{\substack{\<\l, a+1\>}}-\bigg(\ch^{(b-2,b)}_{a,b,b-1}(\l)-\ch^{(b-2,b+1)}_{a,b,b-1}(\l)\bigg)\bigg|_{\substack{\<\l, b-1\>}}\bigg\}\\
\end{eqnarray}
We now evaluate the integral. As discussed in appendix \ref{Holomorphic}, writing the above integrand as a total derivative will always introduce a factor of $[\l|P|\a\>$. With the choice of $|\a\>=|a\>$, terms that are evaluated on the pole $1/\<\l a\pm1\>$ vanishes since $[a\pm 1|P_{a,a\pm 1}|a\>=0$. Thus the cancellation of CCP and the judicious choice or reference spinor reduces the sum of the bubble coefficient for $n$-point MHV amplitude to simply:
\begin{eqnarray}
\nonumber \cc_{2}(l_1^+,l_2^-)&=&\frac{A_n^{\rm tree}}{\<a,b\>^4}\frac{1}{2\pi i}\int_{\bar\lambda=\tilde\lambda}d{\rm LIPS}\;\;\bigg\{ \bigg( \cg^{(b-1,b+1)}_{a,b,b-1}(\l)-\cg^{(b-1,b+1)}_{a,b,b}(\l)\bigg) \frac{\<b+1, a\>}{\<\l, b+1\>}\< a, \l \>^{3}\bigg|_{\<\l, b+1\>}\\
\nonumber&~&\quad\quad\quad\quad\quad\quad\quad\quad\quad+\bigg(\ch^{(b-2,b)}_{a,b,b-1}(\l)-\ch^{(b-2,b+1)}_{a,b,b-1}(\l)\bigg)\bigg|_{\<\l, b-1\>}\bigg\}\, . \nonumber\\
\end{eqnarray}
Expanding the parenthesis, there are four different terms to be evaluated. Explicit evaluation shows the first two terms sum to cancel the last term. Thus we have:
\begin{equation}\label{Target}
\cc_{2}(l_1^+,l_2^-)=\frac{A_n^{\rm tree}}{\<a,b\>^4}\frac{1}{2\pi i}\int_{\bar\lambda=\tilde\lambda}d{\rm LIPS}\;\;\bigg(\ch^{(b-2,b)}_{a,b,b-1}(\l)\bigg)\bigg|_{\l \to b-1}=\frac{11}{6}A_n^{\rm tree}
\end{equation}
Adding this with the same calculation for the other helicity configuration, $(l_1^-,l_2^+)$, one obtains the desired result, $ \cc_2=\frac{11}{3}A^\text{tree}=-\b_0 A^\text{tree}$


\end{document}